\input psfig.tex
\documentclass{aa}
\usepackage{graphicx}
\begin{document}

\title{The ESO Nearby Abell Cluster Survey
       \thanks{Based on observations collected at the European Southern
               Observatory (La Silla, Chile)}
       \thanks{http://www.astrsp-mrs.fr/www/enacs.html}
      }

\subtitle{XI. Segregation of cluster galaxies and subclustering}
\author{A.~Biviano \inst{1} 
   \and P.~Katgert \inst{2}
   \and T.~Thomas \inst{2}
   \and C.~Adami \inst{3} }

\institute{INAF, Osservatorio Astronomico di Trieste, Italy \and
	    Sterrewacht Leiden, The Netherlands \and
   	    Laboratoire d'Astrophysique de Marseille, France}
\offprints{A.~Biviano, biviano@ts.astro.it}
\date{Received / Accepted}

\titlerunning{ENACS XI: Segregation and subclustering}
\authorrunning{Biviano et al.}

\abstract{
We study luminosity and morphology segregation of cluster galaxies in
an ensemble cluster built from 59 rich, nearby galaxy clusters
observed in the ESO Nearby Cluster Survey (ENACS). The ensemble
cluster contains 3056 member galaxies with positions, velocities and
magnitudes; 96\% of these also have galaxy types. From positions and
velocities we identify galaxies within substructures, viz. as members
of groups that are significantly colder than their parent cluster, or
whose average velocity differs significantly from the mean. \\ We
compare distributions of projected clustercentric distance $R$ and
relative line-of-sight velocity $v$, of galaxy subsamples drawn from
the ensemble cluster, to study various kinds of segregation, the
significance of which is obtained from a 2-dimensional
Kolmogorov-Smirnov test. We find that luminosity segregation is
evident only for the ellipticals that are outside (i.e. not in)
substructure and which are brighter than $M_R = -22.0\pm 0.1$. This is
mainly due to the brightest cluster members at rest at the centre of
the cluster potential. \\ We confirm the well-known segregation of
early- and late-type galaxies. For the galaxies with $M_R > -22.0$ of
{\em all} types (E, S0, S and emission-line galaxies, or ELG, for
short), we find that those within substructure have
$(R,v)$-distributions that differ from those of the galaxies that are
not in substructure. The early and late spirals (Sa--Sb and Sbc--Ir
respectively) that are not in substructure also appear to have
different $(R,v)$-distributions. For these reasons we have studied the
segregation properties of 10 galaxy subsamples: viz. E, S0, S$_e$,
S$_l$ and ELG, both within substructure and outside substructure. \\
Among the 5 samples of galaxies that are {\em not in substructure}, at
least 3 ensembles can and must be distinguished; these are: [E+S0],
S$_e$, and [S$_l$+ELG]. The [E+S0] ensemble is most centrally
concentrated and has a fairly low velocity dispersion that hardly
varies with radius. The [S$_l$+ELG] ensemble is least concentrated and
has the highest velocity dispersion, which increases significantly
towards the centre. The class of the S$_e$ galaxies is intermediate to
the two ensembles. Its velocity dispersion is very similar to that of
the [E+S0] galaxies in the outer regions but increases towards the
centre. \\ The galaxies {\em within substructure} do not all have
identical $(R,v)$-distributions; we need to distinguish at least two
ensembles, because the S0 and [S$_l$+ELG] galaxies have different
distributions in $R$ as well as in $v$. The [S$_l$+ELG] galaxies are
less centrally concentrated and, in the inner region, their velocity
dispersion is higher than that of the S0 galaxies. Our data allow the
other 3 galaxy classes to be combined with these two classes in 4
ways. \\ We discuss briefly how our data provide observational
constraints for several processes inside clusters, like the
destruction of substructure, the destruction of late spirals and the
transformation of early spirals into S0's.
\keywords{Galaxies: clusters: general -- Galaxies: elliptical and 
lenticular, cD -- Galaxies: evolution -- Galaxies: kinematics and 
dynamics -- Cosmology: observations} }

\maketitle

\section{Introduction}
\label{s-intro}

It has been known for a long time that in clusters, galaxies of
different classes have different projected distributions. Oemler
(\cite{oe74}), Melnick \& Sargent (\cite{ms77}) and Dressler
(\cite{dr80a}) were the first to quantify these differences. Dressler
(\cite{dr80a}) showed that the different distributions arise mainly
from the so-called morphology-density relation (MDR): i.e., the
relative fractions of ellipticals, S0's and spirals correlate very
well with local surface density. Hence, the composition of the galaxy
population changes with distance from the cluster centre.

Postman \& Geller (\cite{pg84}) derived an MDR over 6 decades of local
space density from the CfA redshift survey, and in the Pisces-Perseus
supercluster -- and in particular in its long filament -- Giovanelli
et al. (\cite{gi86}) found a clear MDR. In this supercluster, even
early and late spirals have different distributions, and this was also
found for spirals in groups of galaxies (Giuricin et al.
\cite{gi88}). The MDR was also studied in several individual
nearby clusters (e.g.  Andreon \cite{an94}, \cite{an96}; Caon \&
Einasto \cite{ce95}) and in general redshift surveys (e.g. Santiago \&
Strauss \cite{ss92}).

In spite of the wealth of observational data, it is still not totally
clear how the MDR arises. In clusters, galaxy encounters must play a
r\^ole, so that gas-rich disk galaxies cannot survive in the dense
cores of clusters. Contrary to Dressler (\cite{dr80a}), Whitmore \&
Gilmore (\cite{wg91}) and Whitmore et al. (\cite{wi93}) found that
morphological fraction correlates as tightly with clustercentric
distance as with projected density. An explanation for the MDR in a
cold dark matter-dominated universe was given by Evrard et
al. (\cite{ev90}).

The study of the MDR was extended towards higher redshifts, e.g. by
Dressler et al (\cite{dr97}), Couch et al. (\cite{co98}) and Fasano et
al. (\cite{fa00}), and was linked to the more general question of the
evolution of galaxies in environments of different densities (e.g. by
Menanteau et al. \cite{me99}). Dressler et al. (\cite{dr97}) found
that the regular, centrally concentrated clusters at a redshift of
about 0.5 show a strong MDR, as do the low-redshift clusters. However,
the less concentrated and irregular clusters at $z \approx$ 0.5 do not
show a clear MDR, unlike their low-redshift counterparts.

Dressler et al. also noted that the fraction of S0's appears to
decrease quite strongly with increasing redshift (by as much as a
factor of 3 from $z = 0$ to $z \approx 0.5$), and Fasano et al.
(\cite{fa00}) studied this effect in clusters at redshifts between 0.1
and 0.25.  The reality of this decrease was questioned by Andreon
(\cite{an98}) who argued that it is not trivial to establish a
reliable elliptical/S0-ratio.  Also, the elliptical/S0-ratio may not
be very meaningful (even if it can be established accurately) because
the differences between ellipticals and S0's may not be major
(e.g. J{\o}rgensen \& Franx \cite{jf94}).

Morphological segregation in position is often accompanied by
morphological segregation in velocity space, i.e.  galaxies of
different types have different velocity dispersions, or velocity
dispersion profiles (e.g. Tammann \cite{ta72}; Moss \& Dickens
\cite{md77}; Sodr\'e et al. \cite{so89}; Biviano et al. \cite{bi92}). 
The effect is sometimes reported as a correlation between kinematics
and colours (e.g. Colless \& Dunn \cite{cd96}; Carlberg et al.
\cite{ca97a}).

Luminosity segregation was detected by Rood \& Turnrose (\cite{rt68}),
Capelato et al. (\cite{ca80}), Yepes et al. (\cite{ye91}) and
Kashikawa et al. (\cite{ks98}).  Luminosity segregation was detected
both as a segregation in clustercentric distance, and as a kinematical
segregation, viz. the most luminous galaxies have the smallest
velocity dispersion (see e.g. Rood et al. \cite{ro72}). Yet,
kinematical segregation appears to occur mostly (Biviano et
al. \cite{bi92}), if not exclusively (Stein \cite{st97}), for
ellipticals, and much less -- if at all -- for the other galaxy
types. Fusco-Femiano \& Menci (\cite{fm98}) explained the observed
degrees of luminosity segregation by their merging models.

Adami et al. (\cite{ad98a}) studied a sample of about 2000 galaxies in
40 nearby Abell clusters and confirmed that the overall velocity
dispersion depends on galaxy type, and increases along the Hubble
sequence. The velocity dispersion profiles for the various galaxy
types indicate that the spirals may not yet be fully virialized, and
may still be mostly on radial, infalling orbits. The spirals may thus
have properties similar to the galaxies with emission lines (ELG),
which were studied in A576 by Mohr et al. (\cite{mo96}), and by
Biviano et al. (\cite{bi97}, hereafter Paper III) in the clusters
observed in the ESO Nearby Abell Cluster Survey (ENACS). Biviano et
al. concluded that the ELG probably have a significant velocity
anisotropy. De Theije \& Katgert (\cite{tk99}, hereafter Paper VI)
distinguished early- and late-type galaxies in the ENACS from their
spectra, and concluded that the evidence for radial orbits was only
significant for the ELG.

Recently, Thomas \& Katgert (\cite{tk02}, hereafter Paper VIII)
derived morphologies for close to 2300 ENACS galaxies from CCD
imaging. By adding morphologies from the literature, and spectral
types from the ENACS spectra, this provides essentially complete type
information for the galaxies in a sample of 59 ENACS clusters. This
dataset allows a vastly improved analysis of the distribution and
kinematics of the various classes of galaxies in clusters, which we
present in this paper. In a subsequent paper (Katgert et al.
\cite{ka02}) we derive the mass profile in the ENACS clusters.

In Sect.~\ref{s-data} we summarize the data that we used. In
Sect.~\ref{s-segr} we discuss the method by which we study the various
types of segregation. In Sect~\ref{s-subs} we discuss the effect of
substructure and in Sects.~\ref{s-lums} and ~\ref{s-morphs} we discuss
the evidence for luminosity and morphology segregation, as well as the
minimum number of galaxy ensembles that must be distinguished. In
Sect.~\ref{s-nat_seg} we discuss the nature of the morphological
segregations and in Sect.~\ref{s-disc} we discuss the implications of
our results for ideas about cluster galaxy evolution. In
Sect.~\ref{s-summ} we present a summary and the main conclusions.

\section{The data}
\label{s-data}

We use data from the ENACS (see Katgert et al. \cite{ka96} -- Paper I
-- and Katgert et al. \cite{ka98} -- Paper V). We have imposed a
redshift limit of $z < 0.1$ and we applied a lower limit of 20 to the
number of member galaxies; this defines a sample of 67 clusters that
is essentially volume-limited (Mazure et al. \cite{ka96} -- Paper II).
Clusters were defined in redshift space, from the distribution of the
line-of-sight velocities. We used a density-dependent gap (Adami et
al. \cite{ad98b}) rather than a fixed gap (as was used in Paper I) to
accommodate different total numbers of galaxies. The membership of the
ENACS clusters with at least 20 redshifts hardly changes when we use a
variable instead of a fixed gap.

Interlopers (non-members, seen in projection onto the cluster) were
eliminated with the interloper removal procedure devised by den Hartog
\& Katgert (\cite{hk96}). For the systems with at least 45 galaxies
with redshifts ($N_z \ge 45$) we calculated an 'interim' mass profile.
This predicts the maximum line-of-sight velocity at the projected
position of each galaxy from which we determine if the galaxy can be
within the turn-around radius. This procedure was repeated until it
converged. For clusters with $N_z \la 45$, such a procedure generally
does not work; therefore we used for them the separation between
members and interlopers as defined in a statistical manner by the $N_z
\ge 45$ clusters (for details, see Katgert et al. \cite{ka02}).

Our magnitudes are R-band, and the absolute magnitudes, $M_R$, were
derived for H$_0$ = 100 km~s$^{-1}$~Mpc$^{-1}$, with K-corrections
according to Sandage (\cite{sa73}) and correction for galactic
absorption according to Burstein \& Heiles (\cite{bh82}). For the
galaxies in our sample, these corrections are quite small, viz. $\sim
0.1$ mag. Information on galaxy type comes from various sources:
either from a CCD image (mostly from Paper VIII), or from the ENACS
spectrum (using a Principal Component Analysis in combination with an
Artificial Neural Network, see Paper VI). A comparison of the various
type estimates and a discussion of their robustness is given in Paper
VIII. Several hundred galaxies have one or more emission lines in
their ENACS spectrum (we refer to those as ELG, see Paper III). Most
of these ELG have narrow lines due to warm gas.

The galaxies with type information were assigned to the following
classes: ellipticals (E), S0, spirals (S), and the intermediate
classes, E/S0, and S0/S. Whenever possible, we also distinguished
between early (S$_e$) and late (S$_l$) spirals, i.e., spirals with
type earlier than or as early as Sb, and later than Sb, respectively.
The classes E, S0, S and ELG are `pure' and `exclusive': i.e. the E,
S0 and S do not contain ELG; as a matter of fact we ignored the ELG in
E's and S0's. The class E/S0 is not used separately, but it is
included in the class of early-type galaxies, together with E and S0,
when these two classes are linked in one sample (see
Sect.~\ref{s-morphs}).  The S0/S class was never used. In clusters
where galaxy types could only be estimated from spectra, the pure E
class does not occur and the `earliest' galaxy class is E/S0 (see also
Paper VIII). Similarly, early and late spiral galaxies can only be
classified on CCD images. However, late spirals can be recognized from
the spectrum alone (see Paper VIII).

We considered including galaxies and clusters with non-ENACS data, but
segregation can only be studied usefully for data with a sufficiently
uniform completeness limit in apparent magnitude. For literature data
this requirement often is not met, so {\em literature data were not
used to enlarge the ENACS galaxy samples}; we only used {\em galaxy
types} from the literature if there was no ENACS galaxy type.
Redshifts from the literature were only used in the identification of
interlopers in the ENACS galaxy samples.

The analysis of the distribution and kinematics of the various galaxy
classes can only be done for clusters with galaxy types for a
sufficiently high fraction of the galaxies, and we required this
fraction to be at least 0.80. This defines a sample of 59 ENACS
clusters with $z < 0.1$; all clusters have 20 or more ENACS member
galaxies, for at least 80\% of which a galaxy type is known. The total
number of galaxies in the 59 clusters is 3056, and for 2948 (96\%) of
those a galaxy type is known. In the sample of 59 clusters, there are
429 ELG, i.e. galaxies with one or more emission lines in the
spectrum. Information about the 59 clusters is given in
Appendix~\ref{s-samples}.

\begin{figure}
\resizebox{\hsize}{!}{\includegraphics{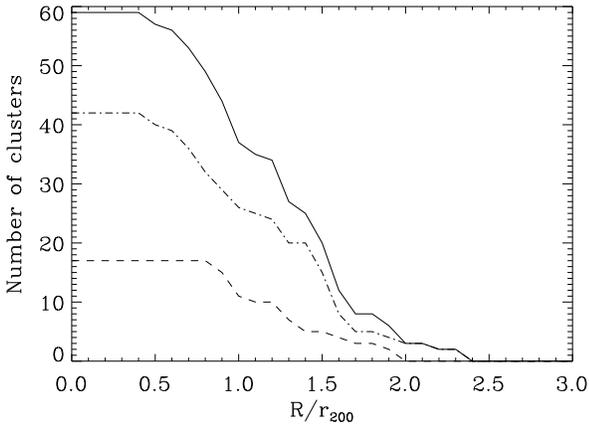}}
\caption{The number of clusters contributing to the analysis, as a 
function of $R/r_{200}$, for the various galaxy classes. The
full-drawn line refers to all 59 clusters, the dashed line to the 17
clusters with only spectral galaxy types (i.e. no pure E-type nor
S$_e$), and the dot-dashed line to the 42 clusters with CCD imaging
(i.e. with all galaxy types)}
\label{f-ncls}
\end{figure}

Combination of data in clusters of various sizes and masses into an
ensemble cluster requires that projected distances and relative
velocities (or rather, their line-of-sight components) are properly
scaled. Projected distances $R$ were scaled with $r_{200}$, the radius
within which the average density is 200 times the critical density of
the universe and which is very close to the virial radius. We assumed,
like Carlberg et al. (\cite{ca97b}), that $M(<r) \propto r$, so that
$r_{200}$ follows from the global value of the dispersion of the
line-of-sight component of the velocities, $\sigma_p$; viz. $r_{200}
\approx \sqrt{3} \, \sigma_p/(10 \, \mathrm{H}(z))$, with $H(z)$ the 
Hubble parameter at redshift $z$. The line-of-sight components
$(v-\overline{v})$ of the relative velocities of the galaxies were
scaled with the global velocity dispersions $\sigma_p$ of the parent
cluster. Note that $\sigma_p$ was calculated for all galaxies
together, irrespective of type, so that the relative velocities of the
{\em different types of galaxies} are all normalized by the {\em same
overall velocity dispersion}. For the clusters in
Table~\ref{t-systems} the average value of $\sigma_p$ is about 700
km/s, so that the average value of $r_{200}$ is about 1.2 h$^{-1}$
Mpc.

In Fig.~\ref{f-ncls} we show the number of clusters contributing in
the various $R/r_{200}$-intervals (solid line). Also shown are the
numbers for the clusters with only spectral galaxy types (i.e. that do
not have E- and S$_e$-types), and clusters with CCD-imaging (that have
all types).

\section{Detecting segregation}
\label{s-segr}

Segregation, i.e., the fact that the various galaxy populations have
different phase-space distributions, may show up either in the
projected distribution, or in the kinematics, or in both. In order to
use our data optimally, we searched for the various types of
segregation by comparing $(R,v)$-distributions in an unbinned way,
using the combined evidence from projected positions and relative
velocities. Luminosity and morphology segregation are generally
presented through compression of $(R,v)$-distributions, i.e., through
projection onto the $R$- or the $v$-axis. Luminosity and morphology
segregation are thus often referred to as 'kinematical segregation'
with respect to magnitude or morphology. However, it is important to
consider radial and kinematical segregation together, because they may
not be independent.

We use an ensemble cluster constructed from the 59 ENACS clusters
listed in Table~\ref{t-systems}.  From this ensemble cluster we select
various galaxy subsamples. Combination of the 59 clusters is necessary
to have the best `signal/noise'. However, the implicit assumption is
that the distributions of the various galaxy type and magnitude
subsamples are sufficiently similar in the individual clusters, or in
different classes of clusters, so that their combination is
meaningful. This is not guaranteed, as cosmic variance is not
negligible. It is therefore possible that no real cluster is described
satisfactorily by the ensemble cluster, but, at the same time, the
ensemble cluster gives the best picture that we have of an average
rich nearby cluster.

For the estimate of the projected clustercentric distance $R$, it is
important that the centre of the parent cluster is as unbiased as
possible. We have taken special care that the centres of all clusters
are determined with similar methods, and with sufficient accuracy. For
the calculation of the cente position we followed the procedure
described in Paper III: in order of decreasing preference we used the
X-ray centre, the brightest cluster member in the core of the cluster,
the peak in the galaxy surface density (if necessary
luminosity-weighted) or the biweight average (e.g., Beers et al.
\cite{be90}) of all galaxy positions to derive the central position.
The estimated accuracy that can be obtained in this way is 50--60 kpc
(see also Adami et al. \cite{ad98c}). The positions of the adopted
cluster centres are given in Table\ref{t-systems} in
Appendix~\ref{s-samples}.

The advantage of comparing $(R,v)$-distributions is that structure in
the $(R,v)$-distribution is not diluted in projection onto either the
$R$- or $v$-axis. However, as a result of the generally non-circular
shapes of the apertures in which the ENACS redshift surveys of the
clusters were done (see, e.g., Fig.~10 in Paper I), the distribution
in projected radial distance is always biased. We estimate this radial
bias by assuming circular symmetry for the cluster galaxy
distributions, knowing the positions (and size) of the Optopus plates
that were used to sample the clusters (see Paper I). Another source of
radial bias arises from the fact that we stack clusters which have
been sampled out to different apertures.  Selection of different
morphological types sometime results in the selection of a subsample
of the original 59 clusters (Fig.~\ref{f-ncls}). Different cluster
subsamples have different degrees of incompleteness at a given
radius. We estimate this radial bias by using an approach similar to
that adopted by Merrifield \& Kent (\cite{mk89}).

When comparing two $(R,v)$-distributions one must
either ensure that these biases are identical, or take the differences
in the biases into account before making the comparison. Since galaxy
subsamples can have different radial distributions, the fact that
clusters have been sampled out to different radii may be a
complicating factor. For all KS2D comparisons in the present paper,
the radial biases in the two $(R,v)$-distributions were found to be
either identical, or so similar that a straight comparison was
justified.

The number of clusters that contribute to the ensemble clusters of the
various galaxy types, at various projected radial distances, is shown
in Fig.~\ref{f-ncls}. In order to ensure that an ensemble cluster is
sufficiently representative (i.e. is built from a sufficient number of
clusters) we have always compared $(R,v)$-distributions over the
radial range $0 \le R/r_{200} \le 1.5$. This means that the ensemble
cluster includes at least 13 out of 59 clusters (or 9 out of 45, for
E, and S$_e$).

The actual comparison of two $(R,v)$-distributions was done with the
2-D version of the Kolmogorov-Smirnov test (KS2D, for short), as
described by Peacock (\cite{pe83}) and Fasano \& Franceschini
(\cite{ff87}). The Kolmogorov-Smirnov test is relatively conservative:
if the Kolmogorov-Smirnov test indicates that two distributions have a
high probability of not being drawn from the same parent population,
other tests (e.g.  Rank-Sum tests or Sign Tests) indicate the
same. However, the Kolmogorov-Smirnov test does not always support
differences indicated by other tests.  Therefore, even if the number
of galaxies in one or both of the samples is relatively small, a small
probability for the samples to be drawn from the same population in
general is trustworthy. However, if the difference is not significant,
this {\em does not prove} that the two samples are drawn from the same
parent population, because real differences can be made undetectable
by limited statistics.

We checked the performance of the KS2D test by randomly assigning half
of the galaxies in each cluster to one of two ensemble clusters,
each comprising half the total number of galaxies in the ensemble
cluster built from the 59 clusters. According to the KS2D-test the
probability that the two subsets are drawn from the same parent
distribution is 83~\%. This is fully consistent with the fact that
they were drawn from the same parent distribution and that there are
no differences between them other than statistical fluctuations.

\section{The effect of substructure}
\label{s-subs}

The combination of data for many clusters into an ensemble cluster is
unavoidable, especially because galaxy subsamples in individual
clusters are too small for segregation studies. However, combining
clusters in an ensemble cluster inevitably reduces the relative
amplitude of substructure. Because substructure may play an important
r\^ole in the formation and evolution of clusters, we have estimated
the effect of substructure on the kinematics and distribution of the
various galaxy classes, in two ways. First, we compared clusters with
and without {\em global} substructure, and secondly we compared
galaxies that reside in and outside {\em local} substructure in their
respective clusters.

\subsection{Clusters with and without substructure}
\label{ss-subg}

Because substructure may take different forms, different filters are
required (and have been devised) for its detection. A general problem
for the detection is that observations only provide the 2+1-D
projected version of the 3+3-D phase-space, which reduces the
detectability. We have used a slightly modified version of the test
devised by Dressler \& Schectman (\cite{ds88}). This test is sensitive to
spatially compact subsystems that either have an average velocity that
differs from the cluster mean, or have a velocity dispersion that
differs from the global one, or both. 

For each galaxy we selected the $n_{loc}$ neighbours that are closest
in projection, where $n_{loc}$ was taken to be $\sqrt{N_{mem}}$ (see,
e.g., Bird~\cite{bi94}) with $N_{mem}$ the total number of cluster members
with redshifts. For these $n_{loc}$ neighbours we calculated the
average velocity, $\overline{v}_{loc}$ and the velocity dispersion
$\sigma_{loc}$. From these parameters, we calculated for each galaxy a
quantity $\delta$, designed to indicate groups (of $n_{loc}$ members)
that are `colder' than the cluster and/or have an average velocity
that differs from the global cluster mean.

The parameter $\delta$ was calculated as follows:
\begin{equation}
\delta = \frac{1}{\sigma_p(R)}\sqrt{\frac{n_{loc} \delta_v^2}
{[t_{n_{loc}-1}]^2}+\frac{\delta_{\sigma}^2}
{\left[1-\sqrt{(n_{loc}-1)/\chi^+_{n_{loc}-1}}\right]^2}}
\label{e-delta}
\end{equation}
\noindent with $\delta_v = \mid \overline{v}_{loc}-\overline{v}_{glob} 
\mid$, and $\delta_{\sigma} = \mbox{max}(\sigma_p-\sigma_{loc},0)$, 
where the Student-t and $\chi^2$ distributions are used to calculate
the uncertainty in the velocity and velocity-dispersion differences,
respectively. To suppress noise, we finally calculated $\delta$ for
each galaxy as the average of the $\delta$-values of its $n_{loc}-1$
neighbours. The larger the value of $\delta$ the larger
the probability that the galaxy finds itself in a moving and/or cold
subgroup within its cluster.

\begin{table}
\caption[]{The evidence for substructure in the clusters with at least 
           45 members with ENACS redshifts}
\begin{flushleft}
\begin{tabular}{|rr|rr|rr|rr|}
\hline
 ACO & $\overline{v}_{3K}$ & \multicolumn{2}{c|}{ENACS} & $P_{\Delta}$ &
 $\sigma_p$  \\ 
  & km/s & z & type &  & km/s \\
\hline
 119 & 12997 & 102 &  87 & 0.917 &  720 \\
 168 & 13201 &  76 &  71 & 0.305 &  518 \\
 514 & 21374 &  82 &  74 & 0.684 &  875 \\
 548 & 12400 & 108 & 108 & 0.005 &  710 \\
 548 & 12638 & 120 & 116 & 0.000 &  824 \\
 978 & 16648 &  56 &  52 & 0.081 &  497 \\
2734 & 18217 &  77 &  77 & 0.010 &  579 \\
2819 & 22285 &  49 &  44 & 0.778 &  409 \\
3094 & 20027 &  66 &  64 & 0.000 &  654 \\
3112 & 22417 &  67 &  60 & 0.211 &  954 \\
3122 & 19171 &  89 &  88 & 0.000 &  782 \\
3128 & 17931 & 152 & 152 & 0.000 &  765 \\
3158 & 17698 & 105 & 102 & 0.771 & 1006 \\
3223 & 17970 &  66 &  65 & 0.204 &  597 \\
3341 & 11364 &  63 &  63 & 0.569 &  561 \\
3354 & 17589 &  56 &  56 & 0.008 &  367 \\
3558 & 14571 &  73 &  73 & 0.063 & 1035 \\
3562 & 14633 & 105 & 105 & 0.000 &  903 \\
3651 & 17863 &  78 &  78 & 0.019 &  662 \\
3667 & 16620 & 103 & 102 & 0.151 & 1037 \\
3806 & 22825 &  84 &  83 & 0.600 &  808 \\
3822 & 22606 &  84 &  68 & 0.079 &  971 \\
3825 & 22373 &  59 &  57 & 0.106 &  699 \\
\hline
\end{tabular}					 
\end{flushleft}
\label{t-pds}
\end{table}				 

As shown in Paper III, at least 40--50 galaxies are needed for a
decision about whether a cluster contains significant substructure or
not. In Table~\ref{t-pds} we list the results for the 23 clusters with
at least 45 members. For each cluster a global $\Delta$ parameter was
calculated as the sum of the individual $\delta$'s of all galaxies.
The observed value of $\Delta$ was compared with the 1000 $\Delta$
values obtained in 1000 random azimuthal scramblings of the galaxy
positions of the same cluster. In the scramblings the incomplete
azimuth coverage was taken into account. The fraction of scramblings
with a value of $\Delta$ larger than the observed one is the
probability $P_{\Delta}$ that the observed value is due to noise, and
thus not indicative of real substructure. Thus, a low value of
$P_{\Delta}$ indicates a high probability of significant substructure.

There are 9 clusters with $P_{\Delta} \le 0.05$, i.e. with significant
global substructure, and these contain 851 galaxies. The other 14
clusters, containing 1069 galaxies, have $P_{\Delta} > 0.05$, i.e are
without significant global substructure. The average number of
galaxies in the substructure clusters is higher than it is in the
non-substructure clusters (95 against 76). This is a reminder that
some of the non-substructure clusters may have substructure that was
not detected due to limited statistics. There is no relation between
the presence of substructure and global velocity dispersion: the 9
substructure clusters have an average velocity dispersion of 694 $\pm$
52 km/sec, for the 14 clusters without substructure this is 763 $\pm$
59 km/sec.

A KS2D comparison of the $(R,v)$-distributions of the total galaxy
populations in substructure and non-substructure clusters shows that
the two samples have a probability of $< 0.1$~\% to have been drawn
from the same parent sample. This forces us to analyze the segregation
properties of galaxies within and outside substructures separately.

\subsection{Galaxies in and outside substructure}
\label{ss-subl}

Instead of summing all individual values of $\delta$ to get a measure
of the amount of substructure in the cluster as a whole (as was done
in Sect.~\ref{ss-subg}), one can also use the individual
$\delta$-values to select galaxies in significant local substructure
in their cluster. In other words: whereas in Sect.~\ref{ss-subg} {\em
all} galaxies in a cluster were made to follow the classification of
their cluster, one may also consider all galaxies in significant local
substructure, independent of the classification of the parent cluster.
Even in clusters without significant {\em global} substructure, some
galaxies may be in {\em local} substructures. Similarly, in clusters
with significant substructure, not all galaxies are in local
substructures.

\begin{figure}
\resizebox{\hsize}{!}{\includegraphics{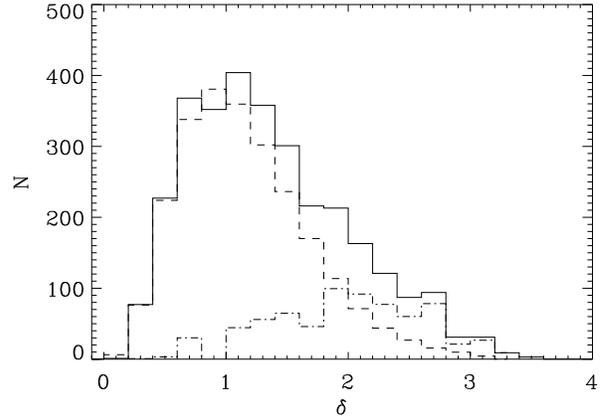}}
\caption{The distribution of $\delta$ for the 3056 galaxies in the 
           59 clusters. The solid line represents the observations,
           the dashed line the distribution for the azimuthally
           scrambled clusters (normalized to the observed number with
           $\delta < 1$), and the dash-dotted line gives the difference
           between the two.}
\label{f-ndelta}
\end{figure}

In Fig.~\ref{f-ndelta} we show the distribution of $\delta$ for all
3056 galaxies in the 59 clusters in our sample. In order to use
$\delta$ to select galaxies in and outside cold and/or moving groups,
we must determine the value of $\delta$ that optimally separates them.
To that end, we azimuthally scrambled the galaxy distributions, taking
into account the incomplete azimuth coverage due to the generally
non-circular shapes of the apertures within which the ENACS
spectroscopy was done. The resulting distribution of $\delta$ (the
dashed line in Fig.~\ref{f-ndelta}) was normalized to produce the
observed numbers of galaxies with $\delta < 1.0$.

This normalization was chosen because for $\delta < 1.0$ no
significant contribution of galaxies in substructure is expected. This
is borne out by the fact that the observed and scrambled $\delta$-
distribution have essentially the same shape for $\delta < 1.0$. The
difference between the observed and 'scrambled' $\delta$ distributions
gives the distribution of the galaxies that presumably are in
substructures (the dash-dotted line in Fig.~\ref{f-ndelta}). The
galaxies that are not in substructures can be selected quite
satisfactorily by requiring $\delta < 1.8$. Both the completeness and
reliability of that sample of 2304 galaxies are very close to 90\%.
We will therefore define all (sub-)samples of galaxies not in
substructure with an upper limit in $\delta$ of 1.8.

The lower limit in $\delta$ for the selection of galaxies within
substructure is less obvious. Using $\delta > 1.8$, the completeness
and reliability of the substructure-sample (of 752 galaxies) are both
about 65\%. I.e., one in three of the galaxies with $\delta > 1.8$ is
{\em not} in substructure. Increasing the lower limit in $\delta$ to
reduce the contamination by galaxies outside substructure also reduces
the number of galaxies in substructure available for the tests. For
comparisons of $(R,v)$-distributions involving samples of galaxies in
substructure, we therefore always defined 4 parallel samples, with
$\delta >$ 1.8, 2.0, 2.2 and 2.4 to vary the balance between
contamination and statistical weight. If the results for those 4
samples are identical, contamination is not important; otherwise the
4 results must be interpreted.

We made a KS2D comparison of the $(R,v)$-distributions of the total
galaxy populations within and outside substructure, using $\delta >$
1.8 for the substructure sample. The probability that the
$(R,v)$-distributions of the two samples are drawn from the same
parent distribution is very small, viz. again $<0.1$~\%.

\subsection{Characteristics of the substructure}
\label{ss-subc}

The definition of substructure that we used was designed to select
cold and/or moving groups. However, because the membership of a group
is fixed to be $n_{loc}$, the properties of the groups are not
necessarily constant. E.g., the 'size' of a group depends on $n_{loc}$
and on the surface density of galaxies, which in turn depends on the
redshift sampling of the cluster $N_{mem}$, so that the groups in
different clusters may have different sizes. This disadvantage is
outweighed by the fact that by setting $n_{loc}$ = $\sqrt{N_{mem}}$
one maximizes the sensitivity to significant substructure while
reducing the sensitivity to Poisson noise (e.g.  Silverman
\cite{si86}). The latter is important, as in our clusters $n_{loc}$
never is as high as the optimum value of $\approx 25$, derived by
Knebe \& M\"uller (\cite{km00}) from an analysis of simulated
clusters.
 
However, even within a cluster the 'size' of the selected groups is
not constant, but varies with distance from the centre because the
surface density increases noticeably towards the center. This effect
is clearly visible in our data: the harmonic mean radius of the
$n_{loc}-1$ neighbours increases with projected distance from the
cluster center. One could avoid this bias by choosing a fixed physical
scale for the selected subclusters. However, if the scale is chosen
large enough for a reasonable number of galaxies to be selected at
large radii, substructure in the central region would be averaged
out. 

Using the $\delta$-values of the individual galaxies, we have
attempted to identify 'subclusters' as follows. First, we selected all
galaxies in each cluster with a value of $\delta > \delta_{lim}$. Then
we calculated the harmonic mean projected radius and the velocity
dispersion of the group of $n_{loc}$ nearest neighbours around each of
these galaxies. These groups were subsequently merged if both their
projected distance was less than the sum of their harmonic mean radii,
and the difference of their average velocities was less than the mean
of their velocity dispersions. The resulting 'subclusters' consist of
all galaxies with $\delta > \delta_{lim}$ in the groups from which
they were built.

For $\delta_{lim} = 2.0$ we find 62 subclusters in the 59 clusters,
i.e. on average one per cluster. The mean number of galaxies with
$\delta > 2.0$ in a subcluster is 8.6, and individual numbers go up to
about 60 (in the very rich cluster A3128). Note that 16 clusters do
{\em not} have a subcluster, while some subclusters are found in
clusters that do not show significant evidence for substructure with
the test of Dressler \& Schectman. Note also that, while the harmonic
mean radius of the selected 'subclusters' is observed to increase with
clustercentric distance, their velocity dispersion stays remarkably
constant, $\sigma_{loc} \sim 400$--500 km~s$^{-1}$.

The filter that we used will not have detected all galaxies that
belong to substructure. Yet, as shown by our azimuthal scramblings, a
large fraction of those that {\em are} selected, {\em do} belong to
subclusters (where this fraction obviously increases with increasing
lower limit in $\delta$). The galaxies selected to be in subclusters
do not form a complete sample, but we treat them as a distinct class,
if only because their $(R,v)$-distribution is likely to be influenced
by the fact that they are dynamically linked in subclusters. This is
actually confirmed by the results of the KS2D tests described in
Sect.~\ref{ss-subl}.

>From histograms like those in Fig.~\ref{f-ndelta} in several radial
intervals, we conclude that the $\delta$-distribution of galaxies in
substructure, corrected for accidental substructure through azimuthal
scrambling (the dashed-dotted line in Fig.~\ref{f-ndelta}), does not
vary significantly with radius. Therefore, we felt justified to apply
radius-independent $\delta$-cuts for galaxies inside and outside of
substructure.

With our normalization of the $\delta$-distribution of the azimuthally
scrambled clusters (see Fig.~\ref{f-ndelta}), the average fraction of
galaxies in substructure, corrected for accidental substructure, is
0.22. However, this fraction appears to depend on the central
concentration of the galaxy distribution. We quantify the latter by a
concentration index, calculated as the ratio of the number of galaxies
within 0.25 $r_{200}$, and between 0.25 and 0.50 $r_{200}$. This index
is not affected by azimuthal incompleteness (Sect.~\ref{s-segr})
because that is negligible within 0.50 $r_{200}$. The 23 clusters
with high concentration index contain 1527 galaxies, the 36
low-concentration clusters contain 1529 galaxies. The corrected
fractions of galaxies in substructure in the two cluster samples are
$0.17\pm0.01$ and $0.27\pm0.01$ respectively. In other words: in
clusters with low central concentration the fraction of galaxies in
substructure is significantly higher than in clusters with high
central concentration. Such a correlation is also seen in numerical
cosmological simulations (Thomas et al. \cite{th01}).

\begin{figure}
\resizebox{\hsize}{!}{\includegraphics{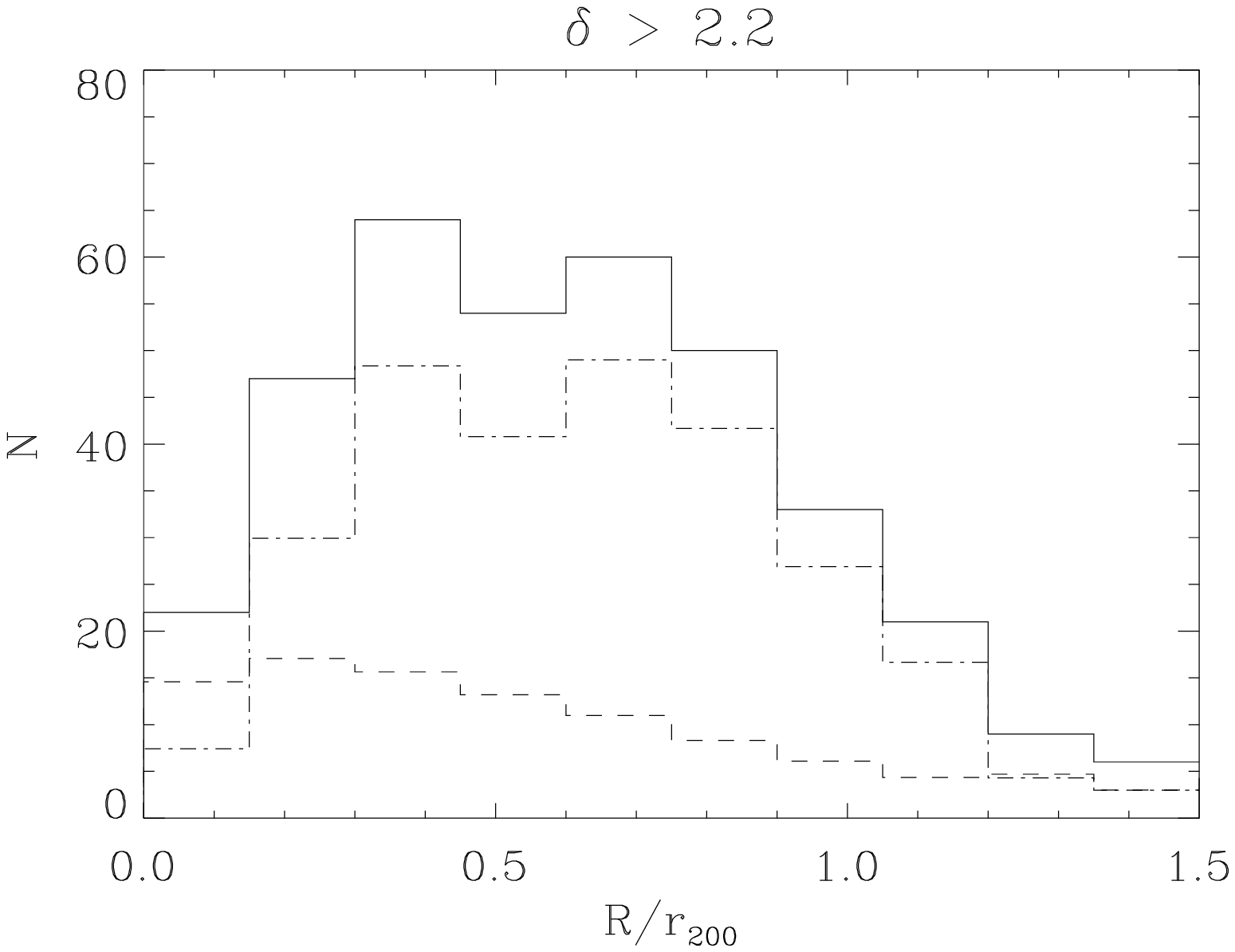}}
\resizebox{\hsize}{!}{\includegraphics{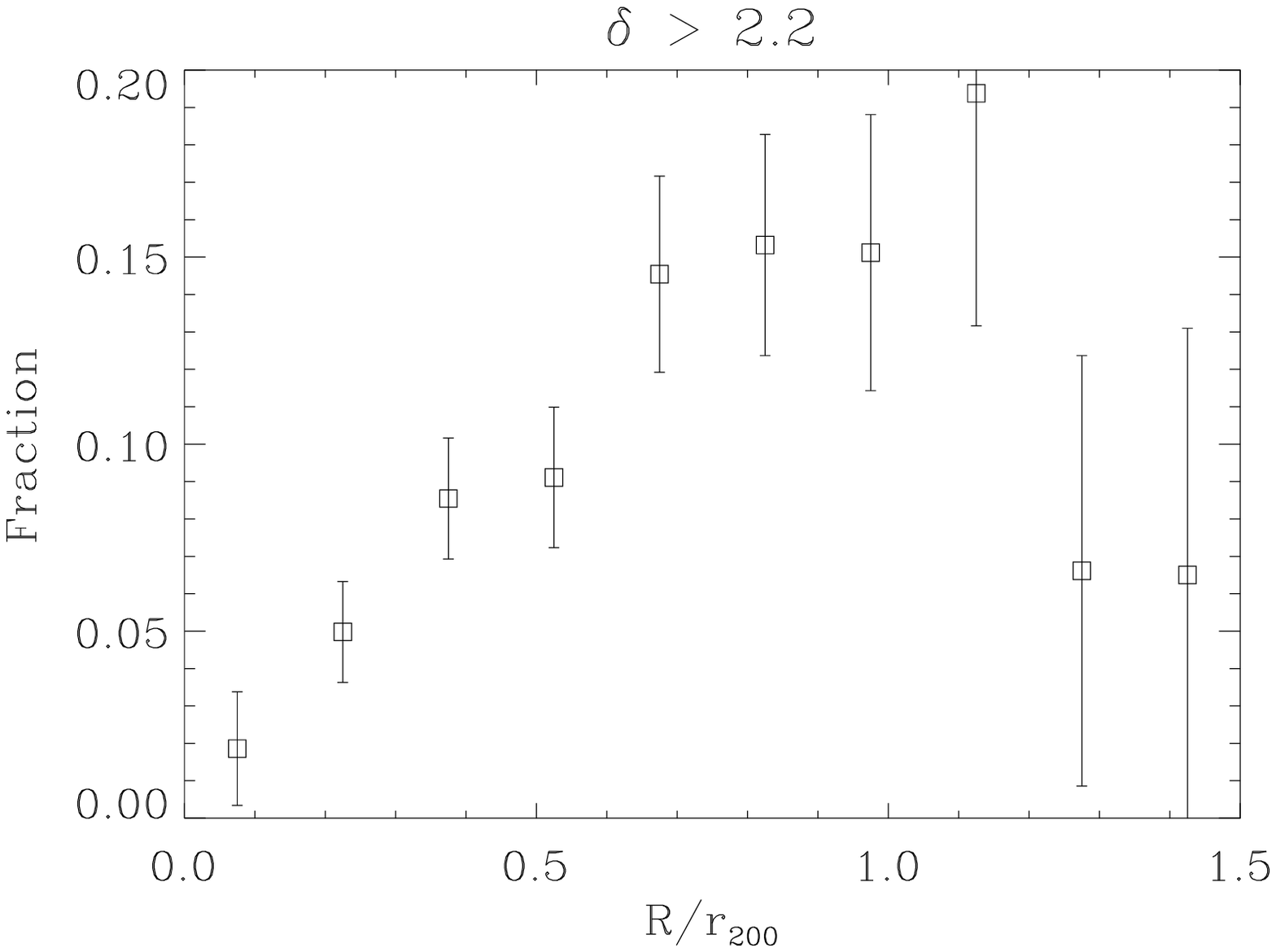}}
\caption{Top: 
           the distribution of the galaxies with $\delta > 2.2$ in the
           59 clusters, as a function of projected radius $R$. The
           solid line represents the observed distribution, the dashed
           line the distribution for the azimuthally scrambled
           clusters, while the dash-dotted line gives the difference
           between the two. Bottom: the fraction of galaxies in
           substructure, with $\delta > 2.2$, as a function of
           projected radius $R$.}
\label{f-ndeltar}
\end{figure}

The number of galaxies in substructure also appears to decrease
markedly towards the centre. This is shown in the upper panel of
Fig.~\ref{f-ndeltar} which gives the radial distribution for the
galaxies with $\delta > 2.2$ (similar results are obtained for other
values of $\delta_{lim}$). The solid line is the observed
distribution, the distribution in the azimuthally scrambled clusters
is given by the dashed line, while the dash-dotted line is the
difference between the two. The latter is a reliable estimate of the
radial distribution of galaxies in substructure. In the lower panel of
Fig.~\ref{f-ndeltar} we show the fraction of galaxies in
substructures, with $\delta > 2.2$, with respect to the total number
of galaxies. The most remarkable feature in Fig.~\ref{f-ndeltar} is
the very strong and abrupt decrease for $R \le 0.3 \, r_{200}$ of the
number of galaxies really in substructure (the dash-dotted line in the
upper panel).

One might wonder to what extent the strong decrease of the number of
galaxies in substructure within $R \approx 0.3 \, r_{200}$ could, at
least partially, be caused by a selection effect. As we discussed
above, the 'size' of the substructure selected by our 'filter' is
smaller in the centre than it is at the periphery. Therefore, we are
insensitive to substructure with a large linear scale in the centre,
and to small-scale substructure in the periphery. Whereas small-scale
substructure might exist in the periphery, it is unlikely that in the
central region large-scale substructure, if it existed, could have
survived. As was shown by Gonz\'{a}lez-Casado et al. (1994), the less
massive subclusters are tidally disrupted in one cluster crossing,
while the more massive clumps migrate towards the centre through
dynamical friction and disappear as substructure. Therefore, we
believe that the strong decrease of the number of galaxies in
substructure towards the centre is real. The effect is reminiscent of
the almost total absence of binary galaxies in the inner region of
rich clusters ($R \la 0.4 h^{-1}$ Mpc), discussed by den Hartog
(\cite{dh97}).

\section{Luminosity segregation}
\label{s-lums}

Evidence for luminosity segregation (LS) is generally presented as a
dependence on magnitude of the distribution of intergalaxy distances,
i.e., of the angular correlation function of the galaxies. The global
character of LS is that the brightest galaxies have a more central
distribution than the other galaxies. As an extreme example, the
brightest galaxies (frequently cD's) are found very close to the
cluster centre. However, not all clusters that have been studied for
LS do show evidence for it.

A robust detection of LS requires a large number of member galaxies.
In several cases, field galaxies are included in the analysis as no
redshifts are available, and those are then a source of noise (see
e.g. Yepes et al. \cite{ye91}; Kashikawa et al. \cite{ks98}). In the
ENACS, field galaxies were eliminated quite well, but the number of
member galaxies in most of the clusters in Table~\ref{t-systems} is
not sufficient to study LS in individual clusters. The ensemble of all
59 clusters does have sufficient statistical weight, but the
combination of many clusters may dilute real LS in (some of) the
individual clusters.

We searched for LS with many KS2D tests in which we compared two
$(R,v)$-distributions of the same class of galaxies, like
ELG$_{nosub}$ or S0$_{sub}$ etc., which differ only in the range of
absolute magnitude. In other words: the parent sample of ELG$_{nosub}$
or S0$_{sub}$ etc. was split in absolute magnitude at $M_{cut}$, the
value of which we varied. For the galaxies in substructure we did the
tests not only for several values of $M_{cut}$ but also for the four
lower limits in $\delta$ that we discussed in Sect.~\ref{ss-subl}. All
those tests show only one robust case of LS: namely for the
ellipticals outside substructure. As elsewhere in this paper,
differences are considered real only if there is less than 5\%
probability that two $(R,v)$-distributions are drawn from the same
parent distribution. For the ellipticals outside substructure, we
consistently get probabilities of less than 5\% for all $M_{cut}$'s in
the range $-22.5$ to $-21.0$.

\begin{figure}
\resizebox{\hsize}{!}{\includegraphics{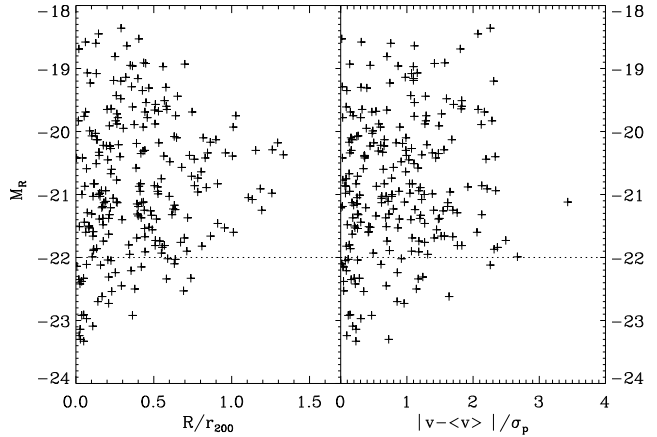}}
\caption{The relations between absolute magnitude and projected radial distance
         (left) and normalized relative velocity
         (right) for ellipticals outside substructure.}
\label{f-lums_e}
\end{figure}

In Fig.~\ref{f-lums_e} we show the relations between absolute
magnitude and projected radial distance (left), and normalized
relative velocity (right), for the ensemble cluster of ellipticals
with $\delta < 1.8$. The brightest ellipticals have velocities close
to the systemic velocity and are mostly in the very centre of their
parent clusters. The fact that the KS2D tests give a signal for a
range of $M_{cut}$ must be due to 'cross-talk': the segregation of the
brightest ellipticals is so strong that it shows up even if fainter
ellipticals are included. To estimate the value of $M_{cut}$ that
optimally separates the bright galaxies that show LS from the faint
ones which do not, we have proceeded as follows.

The faint ellipticals with $-20.6 < M < -20.0$ were taken as a
reference sample, presumably unaffected by LS.  We compared the
$(R,v)$-distribution of this reference sample with that of the
brighter ellipticals with [$M_{cut}-0.25,M_{cut}+0.25]$, for values of
$M_{cut}$ between $-23.25$ and $-21.55$. The faintest $M_{cut}$ for
which the two samples are different is $-22.0$, and we estimate that
the uncertainty in this value is at least 0.1. In the following we
adopt $M_R = -22.0$ as the absolute magnitude above which LS occurs
for ellipticals. Although the KS2D tests do not show significant
evidence for LS of galaxies other than the bright ellipticals, we
decided to exclude galaxies of {\em all} types with $M_R < -22.0$ in
the analysis of morphological segregation, to avoid possible
cross-talk of low-level LS into morphological segregation.

We have investigated the relation between the brightest ellipticals
and the 1st- and 2nd-ranked galaxies in the clusters as follows.
Comparison of the $(R,v)-$distributions of these classes shows that
1st-ranked galaxies (or brightest cluster galaxies, BCG's) and the
brightest ellipticals are not significantly different. On the
contrary, those of the 2nd-ranked galaxies and the brightest
ellipticals {\em are}, and this is also the case for the 1st- and
2nd-ranked galaxies. It is noteworthy that of the brightest
ellipticals 30\% are neither 1st-ranked nor 2nd-ranked galaxies. As a
matter of fact, the average type of 1st-ranked galaxies is
intermediate between E and S0, while that of the 2nd-ranked galaxies
is S0.

\section{Morphology segregation}
\label{s-morphs}

\subsection{Segregation results}
\label{ss-segrules}

We investigated the evidence for morphological segregation by means of
a large number of KS2D comparisons. Because it turned out that the
$(R,v)$-distributions of early and late spirals (S$_e$ and S$_l$) that
are {\em not in substructure} have less than 4\% probability of being
drawn from the same parent distribution, we did not consider them
together. For the spirals {\em within substructure} there is no
evidence that we must consider S$_e$ and S$_l$; however, for
consistency, we also treated them separately. We therefore made KS2D
comparisons of the $(R,v)$-distributions of the following 10 galaxy
classes: E$_{nosub}$, S0$_{nosub}$, S$_{e,nosub}$, S$_{l,nosub}$,
ELG$_{nosub}$, E$_{sub}$, S0$_{sub}$, S$_{e,sub}$, S$_{l,sub}$ and
ELG$_{sub}$. The number of galaxies with $R/r_{200}
\leq 1.5$ in each galaxy class is shown in Table~\ref{t-samples}.

The 5 classes of galaxies {\em outside substructure} were all defined
with a fixed upper limit in $\delta$ of 1.8 (see Sect.~\ref{ss-subl}).
As explained in Sect.~\ref{ss-subl} we defined, for each of the 5
classes of galaxies within substructure, 4 samples with lower limits
in $\delta$ of 1.8, 2.0, 2.2 and 2.4, respectively. From the results
of each set of 4 parallel samples, we gauged the 'diluting' effect (by
contamination of galaxies with a $\delta$-value above the limit, but
which are not in substructure) on real differences (see also
Fig.~\ref{f-ndelta}). At the same time, we estimated the influence of
the opposite effect, viz. a spurious 'difference' due to
contamination. In many cases the results of the 4 samples of galaxies
in substructure with different lower limits in $\delta$ are
consistent. There are 12 comparisons for which there is not total
agreement between the 4 parallel tests. These comparisons are
discussed in more detail in Appendix~\ref{s-ambiv}, where we give the
4 results as well as our interpretation.

With the 10 classes, we did all 45 possible comparisons (which require
a total of 150 KS2D tests, due to the 4 lower limits to $\delta$ used
for the classes of galaxies in substructure). Of these 45 comparisons,
22 show a significant difference. We stress again that the verdict
'significant difference' indicates that the probability that the two
galaxy samples were drawn from the same parent sample is less than
5\%. It should be appreciated that comparisons for which no believable
difference was found are 'undecided'. In other words: in those cases
{\em it is not proven} that the galaxy samples have identical
$(R,v)$-distributions, because our data only indicate that they are
not significantly different.

\begin{table}
\caption[]{The number of galaxies with $R/r_{200} \leq 1.5$ and 
$M_R > -22.0$ in each of the samples used}
\begin{flushleft}
\begin{tabular}{|l|r||r|r|r|r|}
\hline
gal & outside & \multicolumn{4}{c|}{within substructure} \\
\cline{3-6} 
type & substr. & $\delta > 1.8$ & $\delta > 2.0$ & $\delta > 2.2$ & 
  $\delta > 2.4$ \\
\hline
E         & 200 &  60 &  41 &  30 & 15 \\ 
S0        & 795 & 261 & 176 & 114 & 70 \\ 
S$_e$     & 183 &  63 &  41 &  28 & 21 \\ 
S$_l$     & 113 &  25 &  19 &  16 & 12 \\ 
ELG       & 236 &  88 &  73 &  56 & 44 \\
\hline
E/S0      & 165 &  72 &  51 &  36 & 19 \\
S$_{generic}$ & 119 &  33 &  25 &  18 & 13 \\
\hline
\end{tabular}					 
\end{flushleft}
\label{t-samples}
\end{table}				 

The results are as follows:
\begin{itemize}
\vspace*{-0.0cm}
\item
Of the 10 comparisons between classes of galaxies {\bf not in
substructure}, 6 show a difference: viz.  
E--S$_{l}$, \\ E--ELG, S0--S$_{e}$, S0--S$_{l}$, S0--ELG and
S$_{e}$--S$_{l}$.
\item
Of the 10 comparisons between classes of galaxies {\bf in
substructure}, 2 show a difference: viz. S0--S$_{l}$ and \\ 
S0--ELG
\item
Of the 25 'mixed' comparisons between classes of galaxies in and
outside substructure, 14 show a difference. The latter are not very
surprising in view of the result discussed in Sect.~\ref{ss-subl}.
Instead, the comparisons for which {\em no} difference was found may
be more informative in this case; these are: E$_{sub}$--S$_{l,nosub}$,
\\ E$_{sub}$--ELG$_{nosub}$, S0$_{sub}$--S$_{e,nosub}$,
S0$_{sub}$--S$_{l,nosub}$, \\ S0$_{sub}$--ELG$_{nosub}$,
S$_{e,sub}$--E$_{nosub}$, S$_{e,sub}$--S0$_{nosub}$, \\
S$_{e,sub}$--S$_{e,nosub}$, S$_{e,sub}$--S$_{l,nosub}$,
S$_{e,sub}$--ELG$_{nosub}$ and S$_{l,sub}$--S$_{l,nosub}$. \\ In view
of the sample sizes (see Table~\ref{t-samples}) the latter 6 results
may well be due to limited statistics.
\end{itemize}

\begin{figure*}
\resizebox{\hsize}{!}{\includegraphics{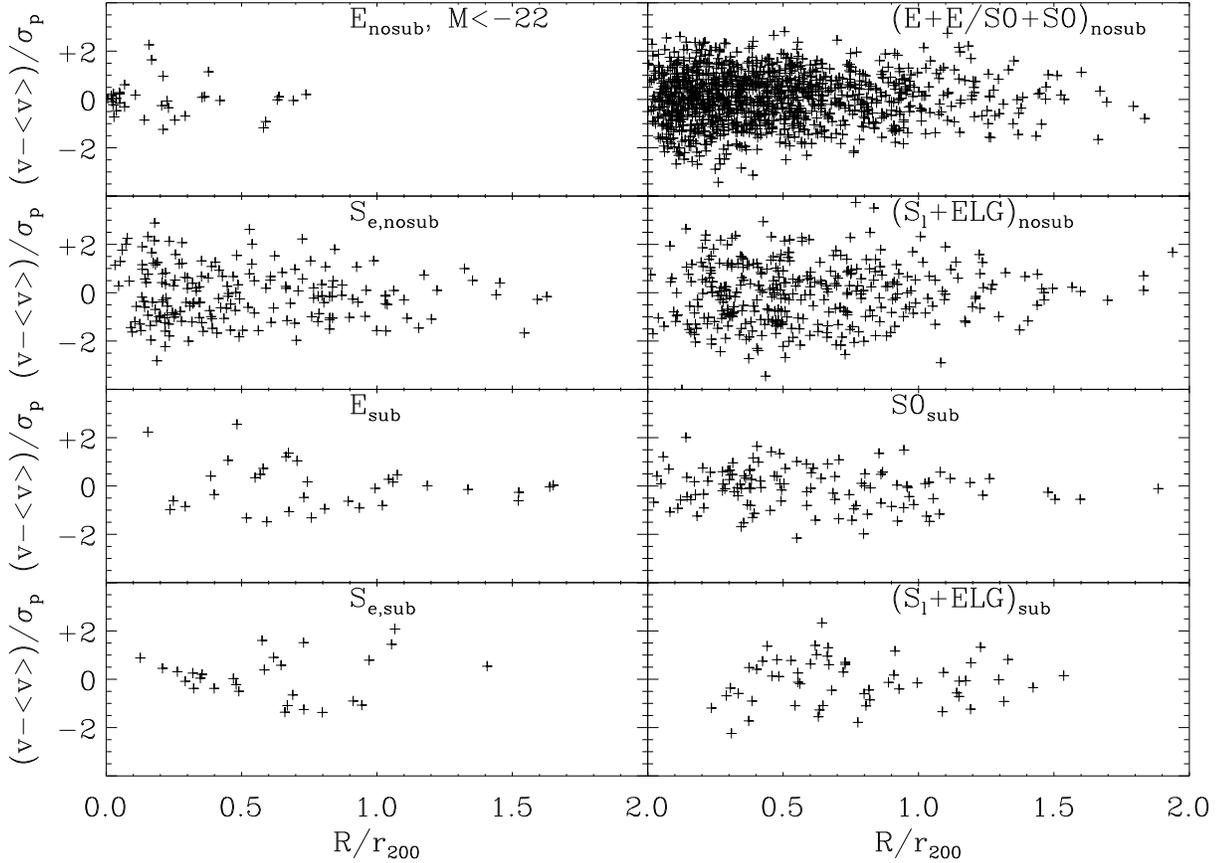}}
\caption{The $(R,v)$-distributions of the 8 classes of 
galaxies implied by the analysis of luminosity and morphology
segregation. The samples in substructure were defined with a
substructure parameter $\delta > 2.2$.}
\label{f-rv8}
\end{figure*}

As mentioned in Sect.~\ref{s-data}, all projected distances $R$ were
expressed in $r_{200}$, which was derived from the velocity dispersion
$\sigma_p$ on the assumption of a mass profile with $M(<r) \propto r$.
We redid all KS2D comparisons with projected distances scaled with an
alternative value of $r_{200}$, calculated for an assumed mass profile
$M(<r) \propto \sqrt{r}$. The results are essentially identical, and
lead to the same definition of galaxy ensembles, as we describe now.

\subsection{The minimum number of galaxy ensembles}
\label{ss-ensembl}

Based on the segregation results discussed in Sect.~\ref{ss-segrules},
we have tried to find the {\em minimum} number of galaxy samples that
{\em must} be distinguished. The KS2D tests tell us which galaxy
classes cannot be combined (i.e. those that have a probability of less
than 5~\% of being drawn from the same parent population), and which
classes {\em can} in principle be combined (those with a probability
of more than 5~\% \ldots), but they do not tell us which ones {\em must}
be combined.

We will first consider the classes of galaxies within and outside
substructure separately. This is motivated by the fact that the
individual galaxy classes, viz. ellipticals, S0's, all spirals
(including the generic spirals) and ELG in and outside substructure
have different $(R,v)$ distributions. Note that this is true for all 4
substructure samples (i.e. for different values of $\delta_{min}$).
It is true that our data do not indicate that S$_{e,nosub}$ and
S$_{e,sub}$ have different $(R,v)$ distributions, and similarly for
S$_{l,nosub}$ and S$_{l,sub}$, but this may be due to limited
statistics.

For the galaxies that are not in substructure, the data do not allow
less than three galaxy ensembles. This is because the S$_e$'s and
S$_l$'s cannot be combined, while at the same time neither of these
can be combined with the S0's. According to the other 'segregation
rules' in Sect.~\ref{ss-segrules}, the E's and ELG can be combined in
three different ways with S0's, S$_e$'s and S$_l$'s. E.g., the E's can
be combined with S0's as well as S$_e$'s, while the ELG can be
combined with S$_e$'s and S$_l$'s, as long as this does not imply
combining them with E's or S0's. Two of these three ensemble
configurations are not meaningful, because in them there are two
ensembles that are not significantly different (i.e, which have more
than 5~\% probability to have been drawn from the same parent
population). That leaves {\em a unique ensemble configuration for the
galaxies that are not in substructure}, viz.:
\begin{itemize}
\item {[E$_{nosub}+$S0$_{nosub}$], S$_{e,nosub}$,
       [S$_{l,nosub}+$ELG$_{nosub}$]}
\end{itemize}
We note in passing that the fact that E and S0 outside
substructures constitute an ensemble means that the intermediate
galaxy class, E/S0$_{nosub}$, can be included in that ensemble.

For the galaxies within substructures, the 'segregation rules' require
a minimum of two ensembles. This is remarkable, because the definition
of substructure and the assignment of galaxies to substructure was
done totally independent of galaxy type. Applying the 'segregation
rules' for the galaxies in substructure, we obtain four possible
two-ensemble configurations, viz.:
\begin{itemize}
\item {S0$_{sub}$, [E$_{sub}+$S$_{e,sub}+$S$_{l,sub}+$ELG$_{sub}$]}  
\item {[S0$_{sub}+$E$_{sub}$],[S$_{e,sub}+$S$_{l,sub}+$ELG$_{sub}$]}  
\item {[S0$_{sub}+$S$_{e,sub}$], [E$_{sub}+$S$_{l,sub}+$ELG$_{sub}$]} 
\item {[S0$_{sub}+$E$_{sub}+$S$_{e,sub}$], [S$_{l,sub}+$ELG$_{sub}$]} 
\end{itemize}
In each of these configurations, the two ensembles always have
significantly different $(R,v)$ distributions. Because there is no
evidence that we need to separate S$_{e,sub}$ and S$_{l,sub}$, we add
the generic spirals (within substructures) to the two ensembles which
contain both S$_{e,sub}$ and S$_{l,sub}$.

While we were able to identify a unique 3-ensemble configuration for
the galaxies outside substructure, our data do not uniquely define the
two ensembles into which the galaxies within substructures are
segregated. Nevertheless, the 10 galaxy classes that we started with
can be reduced to 5 ensembles. However, the choice between the 4
possible 5-ensemble configurations cannot be made with our data.

So far, we have only considered ensembles that consist only of
galaxies either in or outside substructure. Yet, our data allow
combinations of galaxies in and outside substructure in a single
ensemble. If such combinations are not considered unacceptable for
physical reasons, one can construct ensemble configurations with 4
ensembles instead of 5. The reason is that, even though galaxies {\em
of a given class} have different $(R,v)$-distributions within and
outside substructures (except maybe the S$_e$ and S$_l$), this is not
true in general. In other words: the KS2D tests allow e.g. S0$_{sub}$
and ELG$_{nosub}$ to be combined. However, no configurations with 3
ensembles are possible. This is because, according to our data, the
[E+S0]$_{nosub}$ cannot be combined with any of the 8 substructure
ensembles. The fact that the [E+S0]$_{nosub}$ can be combined with the
S$_{e,sub}$ which are in one of the two substructure ensembles does
not affect that conclusion.

If one tries to construct mixed ensembles, simply by combining the 3
ensembles outside substructure with the 4 sets of 2 ensembles within
substructure, one can construct 6 configurations of 4 ensembles.
However, there is no good reason why one could then not 'open' the 3
ensembles outside substructure and the two within substructure, and
the number of possible 4-ensemble configurations then certainly
becomes larger than six. However, we do not consider it very useful to
explore all those possibilities.

\section{The nature of the morphological segregations}
\label{s-nat_seg}

Having studied, through KS2D tests in which we compare
$(R,v)$-distributions, which morphological segregations are indicated
by our data, it remains to characterize the nature of the various
segregations. In Fig.~\ref{f-rv8} we show the $(R,v)$-distributions of
the following 8 galaxy classes: {\em outside substructure,} the
E$_{nosub}$ with $M<-22$, the [E+E/S0+S0]$_{nosub}$, the
S$_{e,nosub}$, the [S$_l$+ELG]$_{nosub}$, and {\em in substructure}
the E$_{sub}$, the S0$_{sub}$, the S$_{e,sub}$ and the
[S$_l$+ELG]$_{sub}$. As explained in Sect.~\ref{s-lums}, the last 7
classes do not contain galaxies with $M<-22$. In order to minimize the
effects of contamination in the substructure samples, while still
retaining a reasonable statistics, we applied a lower limit in
$\delta$ of 2.2 (see also Sect.\ref{ss-subl}). Because ellipticals and
S0's outside substructure have $(R,v)$-distributions that are not
significantly different, we added the class E/S0$_{nosub}$ to that
sample.

Fig.~\ref{f-rv8} visually illustrates the segregation results
discussed in Sects.~\ref{s-lums} and \ref{s-morphs}. The bright
ellipticals clearly have a very distinct $(R,v)$-distribution, unlike
that of any of the other classes, and the physical reason for that has
been amply discussed in the literature (see, e.g., Governato et al.
\cite{go01}). Note that the 4 classes of galaxies outside substructure 
are mutually exclusive, but the 4 classes of galaxies inside
substructure are not, and we could combine them into 4 possible
configurations of 2 ensembles (see Sect.~\ref{ss-ensembl}). However,
we do not show the $(R,v)$-distributions of all these possible
configurations here.

\subsection{The ensembles of galaxies outside substructure}
\label{ss-nat_seg_out}

The 3 ensembles of galaxies that are not in substructure, viz.
[E+E/S0+S0]$_{nosub}$, S$_{e,nosub}$ and [S$_l$+ELG]$_{nosub}$ are
indeed seen to have different $(R,v)$-distributions. The nature of
these differences is illustrated in the form of cumulative
distributions of $R$ and $v$ in Fig.~\ref{f-cumrv_nosub}.
Fig.~\ref{f-cumrv_nosub}a shows clearly that the [E+E/S0+S0]$_{nosub}$
(the early-type galaxies) are the most centrally concentrated of the 3
ensembles. For $R/r_{200}\la 0.5$, the shape of the distribution of
the S$_{e,nosub}$ is quite similar to that of the early-type galaxies,
but at larger distances it may become slightly wider. The distribution
of the [S$_l$+ELG]$_{nosub}$ (the late-type galaxies) is widest of
all, and flattens strongly towards the centre.

Because the velocity distributions depend on $R/r_{200}$, we show, in
Fig.~\ref{f-cumrv_nosub}b, the cumulative velocity distributions for
three radial intervals. In the inner region ($R/r_{200}<0.25$) the
S$_{e,nosub}$ and the late-type galaxies have very similar velocity
distributions, which are significantly broader than that of the
early-type galaxies. Instead, in the intermediate radial interval
($0.25\le R/r_{200}<0.75$), the velocity distributions of the
S$_{e,nosub}$ and early-type galaxies are very similar, while that of
the late-type galaxies is broader than the other two. Finally, for
$0.75\le R/r_{200} \le 1.5$ the three distributions are quite similar,
although there is a weak hint that now the velocity distribution of
the S$_{e,nosub}$ galaxies may be even narrower than that of the other
ensembles. Although the differences between early- and late-type
galaxies are not completely new, they are now demonstrated with
unprecedented statistical weight and detail.

However, the behaviour of the S$_{e,nosub}$ galaxies was not seen
before and is very intriguing. The segregation of S$_{e,nosub}$ and
S$_{l,nosub}$, although significant with our limit of 5\% probability,
is not the strongest segregation that we find. Therefore, it is
gratifying to see that there is also a good physical reason for
distinguishing the S$_{e,nosub}$ as a separate class, namely the fact
that their velocity distribution changes `allegiance' from inside to
outside. Further support for the reality of and need for a separate
S$_{e,nosub}$ class comes from the KS2D comparisons of the
S$_{e,nosub}$ with the [E+E/S0+S0]$_{nosub}$ and the
[S$_l$+ELG]$_{nosub}$, both of which show the three ensembles to be
different.

\begin{figure}
\resizebox{\hsize}{!}{\includegraphics{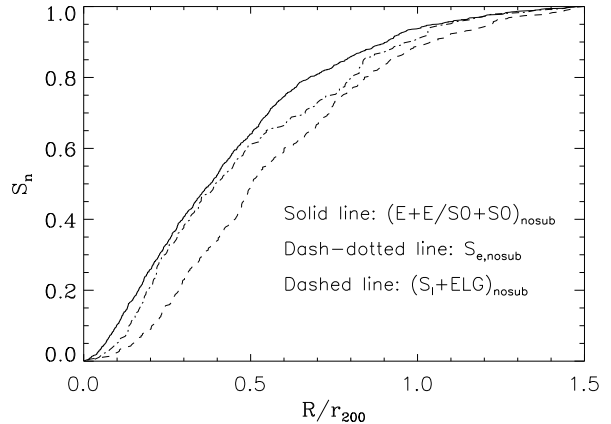}}
\caption{{\bf a.} The cumulative $R$-distribution of the 3 ensembles
outside substructure.}
\end{figure}
\addtocounter{figure}{-1}

\begin{figure}
\resizebox{\hsize}{!}{\includegraphics{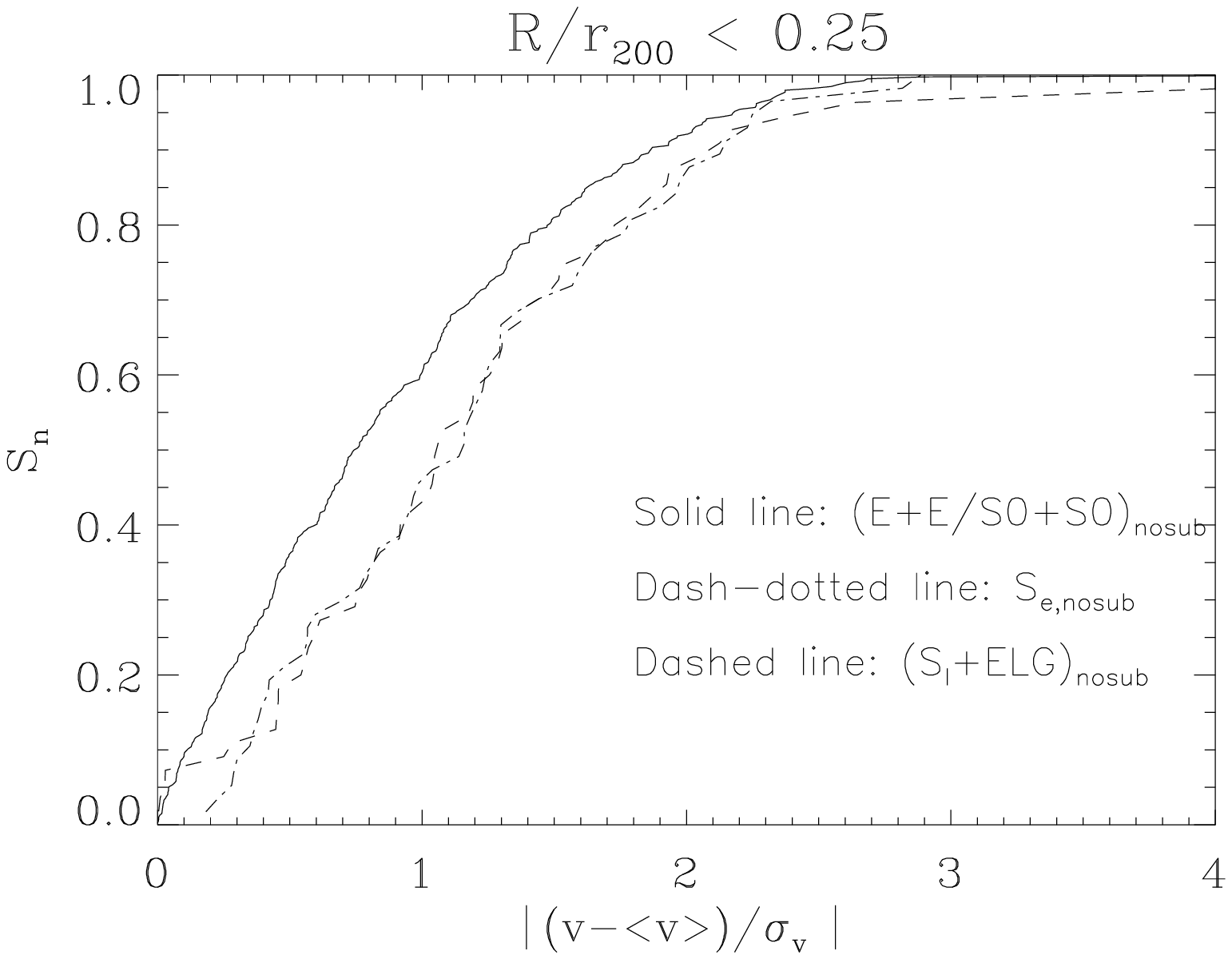}}
\resizebox{\hsize}{!}{\includegraphics{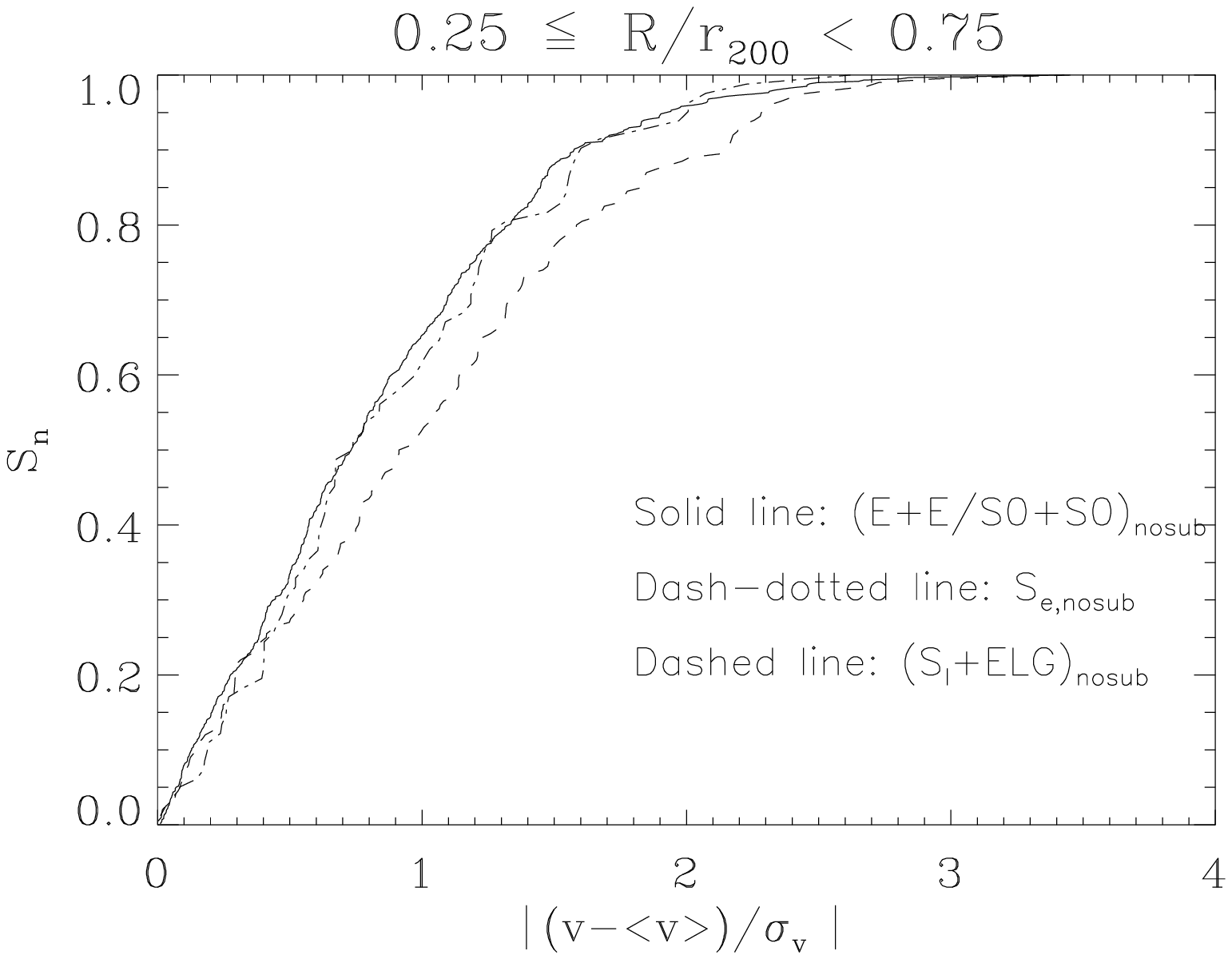}}
\resizebox{\hsize}{!}{\includegraphics{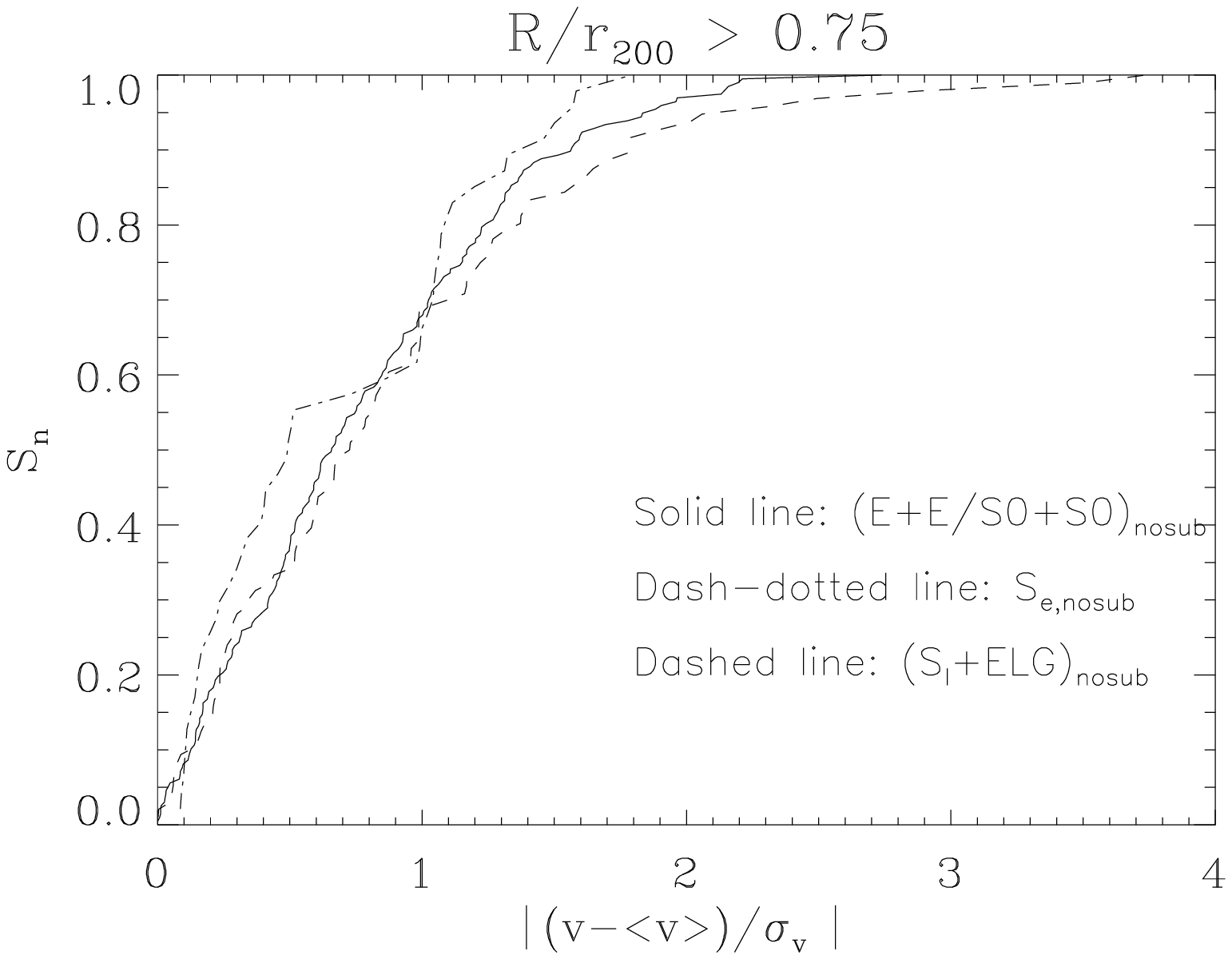}}
\caption{{\bf b.}The cumulative $v$-distributions of the 3 ensembles
outside substructure, in three different radial ranges.}
\label{f-cumrv_nosub}
\end{figure}

\subsection{The ensembles of galaxies in substructure}

The segregation of the classes of galaxies in substructure is much
less clean-cut than it is for those outside substructure. In itself,
it is remarkable that the galaxies inside substructure should show
segregation at all, because the selection of galaxies with $\delta >
\delta_{lim}$ (with $\delta_{lim}$ = 1.8, 2.0, 2.2 and 2.4) was made
independent of galaxy type. Yet, the KS2D tests indicate that the S0's
in substructure have an $(R,v)$-distribution that is different from
both S$_l$ and ELG separately, as well as from [S$_l$+ELG]. The
$(R,v)$-distributions of S0$_{sub}$ and [S$_l$+ELG]$_{sub}$ in
Fig.~\ref{f-rv8} (where $\delta_{lim} = 2.2$) provide visual support
for the difference.

\begin{figure}
\resizebox{\hsize}{!}{\includegraphics{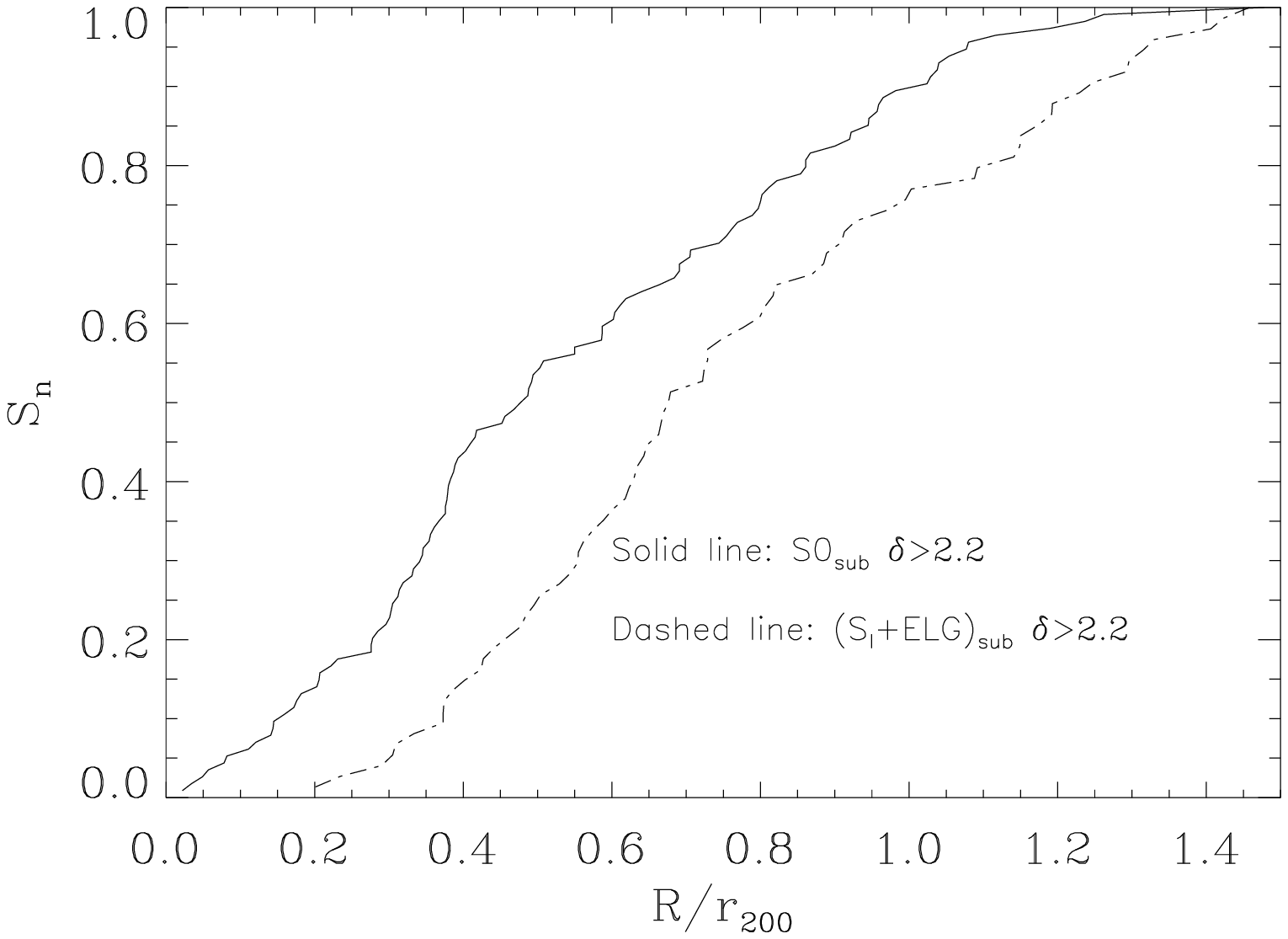}}
\caption{{\bf a.} The cumulative $R$-distributions of S0 and [S$_l$+ELG]
galaxies in substructure.}
\end{figure}
\addtocounter{figure}{-1}

\begin{figure}
\resizebox{\hsize}{!}{\includegraphics{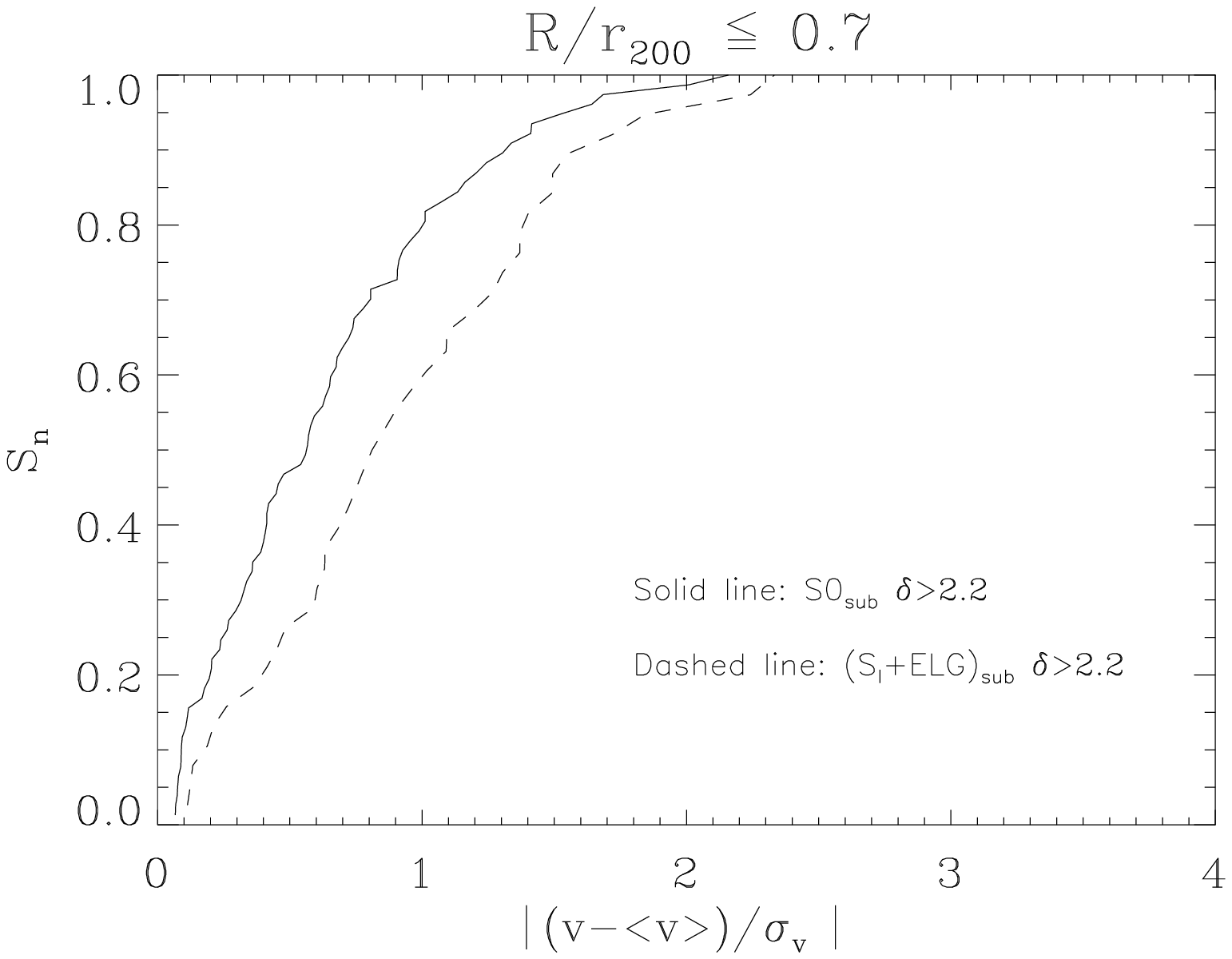}}
\resizebox{\hsize}{!}{\includegraphics{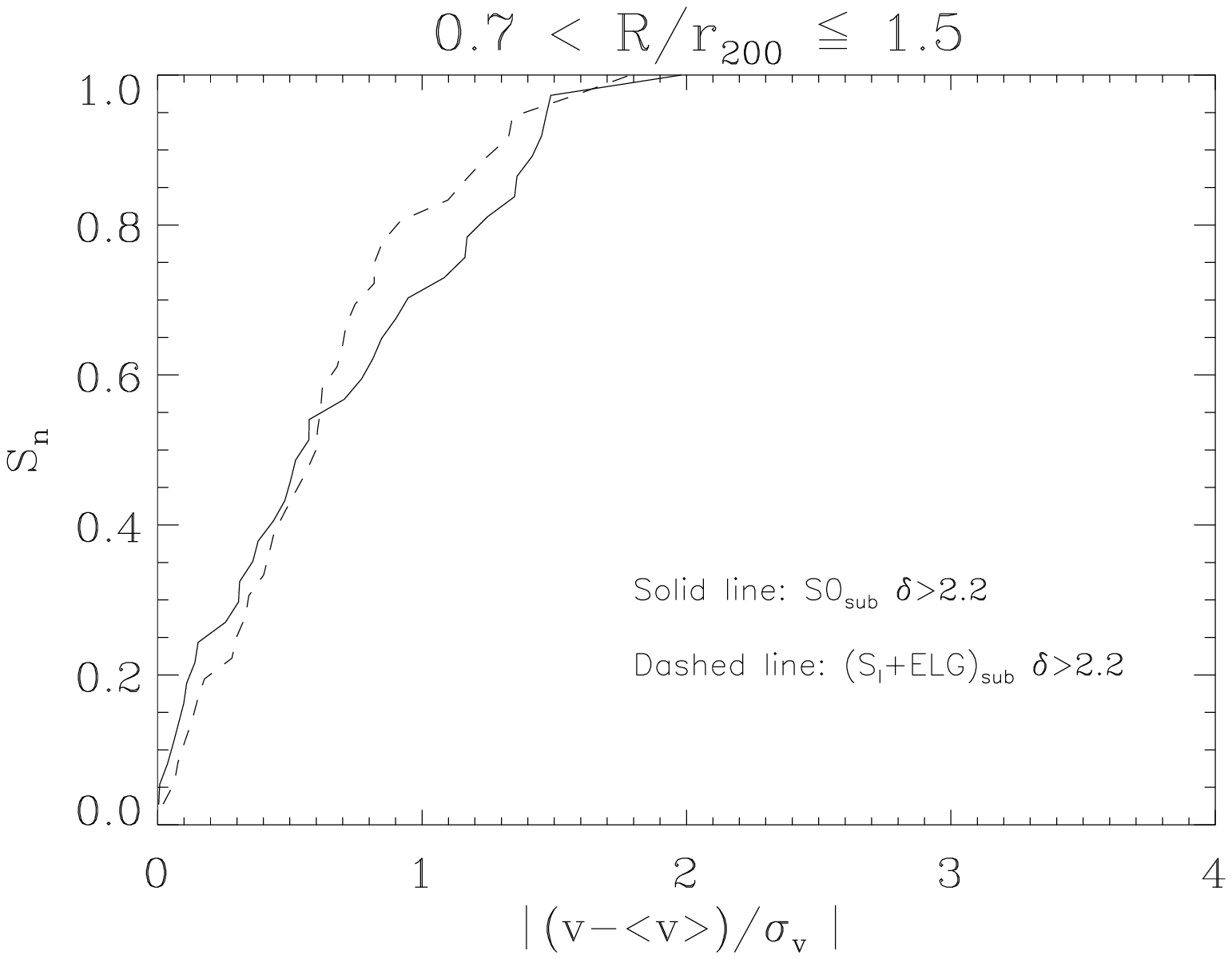}}
\caption{{\bf b.} The cumulative $v$-distributions of S0 and [S$_l$+ELG]
galaxies in substructure, in two different radial ranges.}
\label{f-cumrv_sub}
\end{figure}

As we did for the ensembles outside substructure, we illustrate the
nature of the segregation between S0$_{sub}$ and [S$_l$+ELG]$_{sub}$
with cumulative distributions of projected distance and relative
velocity.  In Fig.~\ref{f-cumrv_sub}a we show the cumulative radial
distributions for $\delta_{lim} = 2.2$. The difference between the two
radial distributions is quite clear. Yet, contamination by S0's
outside substructure could be partly responsible for the difference.
However, the difference in Fig.~\ref{f-cumrv_sub}b indicates that the
late-type galaxies in substructure not only have a shallower radial
distribution but also a larger velocity width than the S0's. This
latter fact cannot be attributed to contamination by S0's outside
substructure, particularly because the latter have (if anything) a
larger velocity width than the S0's inside substructure.

\section{Discussion}
\label{s-disc}

\subsection{Summary of our findings}
\label{s-dis_summ}

We start the discussion of the implications of the results of our
analysis by summarizing those results and by putting them in a broader
perspective.

Our {\em first conclusion} concerns {\em luminosity segregation}. We
only find significant luminosity segregation for the brightest E's
outside substructure. Had the S0's outside substructure shared the
luminosity segregation of the E's, the statistical weight of our data
would have been more than sufficient to detect it. Thus, our result is
more in line with that by Stein (\cite{st97}) than with that by
Biviano et al. (\cite{bi92}). The absolute magnitude $M_R$ of $-22$
which separates the brightest E's (those that show LS) from the other
E's agrees nicely with that found by Biviano et al. (\cite{bi92}) as
the limit for kinematical segregation (using $V-R \simeq 0.6$ for
early-type galaxies, see, e.g., Poulain \& Nieto \cite{pn94}). Of the
42 E's in our sample with $M_R < -22$, about two-thirds are
first-ranked galaxies. The brightest E's, which mostly occur outside
substructure, thus appear to form a really separate population which
is very centrally concentrated and kinematically quite cold, and which
probably mostly has an accretion and merger origin (see, e.g.,
Governato et al. \cite{go01}).

Global estimates of the timescale for dynamical friction in an average
ENACS cluster yields about 1 Gyr for the brightest ellipticals and for
the 1st-ranked galaxies, and about 2 Gyr for the 2nd-ranked galaxies.
So, dynamical friction can be very well responsible for the luminosity
segregation that we observe.  Given these estimates it is perhaps
somewhat surprising that the 2nd-ranked galaxies do not show evidence
for luminosity segregation. This could indicate that the 2nd-ranked
galaxies are being cannibalized by the 1st-ranked galaxies when they
get too close to the cluster centre.

The {\em second conclusion} concerns {\em substructure}. For E's,
S0's, spirals as well as ELG, the $(R,v)$-distributions of galaxies in
and outside substructure are significantly different. This mostly
reflects the fact that galaxies in and outside substructure have very
different radial distributions (see Fig.~\ref{f-rv8}). In particular,
the small fraction of galaxies in substructure within $R
\approx 0.3 \, r_{200}$ probably supplies most of the 'signal' for the
differences detected by the KS2D tests. As we argued in
Sect.~\ref{ss-subc}, this effect is most likely real and not induced
by the radial dependence of the 'size' of groups in our selection of
galaxies in substructure.

The {\em third conclusion} concerns the {\em galaxies outside
substructure}. We find that we must distinguish 4 different classes,
viz: 1) the brightest E's, 2) all but the brightest E's combined with
the S0's, 3) the S$_{late}$ combined with the ELG's and 4) the
S$_{early}$'s. Thus, excluding the brightest E's, the projected
distribution and kinematics of the E's and the S0's are not
significantly different. This would follow naturally if these two
classes had a common origin and evolution, or if they formed one
class. The latter was suggested by J{\o}rgensen \& Franx
(\cite{jf94}), who concluded that the E's and S0's form one class with
a continuous change in L$_{disk}$/L$_{total}$, with the different
classifications mostly induced by the viewing angle. It is true that
structural differences between E's and S0's now appear much smaller
than was once thought, while the stellar populations are also very
similar. However, Thomas \& Katgert (\cite{tk02}) conclude, from
samples of 194 E's and 307 S0's in 42 ENACS clusters that viewing
angle plays an important r\^ole, but is probably not the only factor
that determines the outcome of the classification. It is also not
clear that the morphology-density relation, and its reported
dependence on redshift (e.g. Dressler et al. \cite{dr97}) is
consistent with the picture of J{\o}rgensen \& Franx (\cite{jf94}).

The late spirals and ELG cannot be distinguished either from their
distributions in the $(R,v)$-plane. This is not very surprising as a
large fraction of the ELG (which are all late-type galaxies as the few
ELG associated with early-type galaxies were excluded) is associated
with {\em late} spirals. However, the process responsible for the
removal of the gas that gives rise to the emission lines apparently
does not create any differences between the spatial distribution and
kinematics of the late spirals without gas and those with gas
(i.e. the ELG). As a matter of fact, our result confirms the general
assumption that the gas-removal process changes only the appearance of
the galaxy but not its velocity. In other words: we should have been
very surprised if the observations had shown differences in
$(R,v)-$distributions as a result of the gas-removal process.

Finally, the early spirals appear to be a separate class among the
galaxies outside substructure. This is quite a robust result: the
$(R,v)$-distributions of the 183 S$_e$'s and the 1160 E+E/S0+S0's have
a 4.1\% probability to have been drawn from the same parent
distribution, and for the 183 S$_e$'s and the 349 S$_l$+ELG's this is
2.0\%. This result is also new and it may have important consequences
for the picture of the evolution and transformation of galaxies in
clusters. Most intriguing is the fact that the velocity distribution
of the S$_e$'s is very similar to that of the S$_l$+ELG's in the very
centre, while it is closer to that of the E+S0's beyond $\sim 0.3 \,
r_{200}$.

The {\em fourth conclusion} concerns the {\em galaxies in
substructure}. There are two comparisons that show a significant
difference, viz. S0 vs. S$_l$ and S0 vs. ELG, and since we are allowed
to combine S$_l$ and ELG, we checked that S0 and S$_l$+ELG are also
significantly different. It thus appears that the total fraction of
galaxies in substructure decreases strongly within $R \sim 0.3
r_{200}$, but that the fraction of S0's in substructure increases
within $R \sim 0.3 r_{200}$.

We now turn to a more qualitative discussion of the implications of
our results for current ideas about formation, evolution and
transformation of galaxies in clusters.

\subsection{Galaxies in substructure}
\label{ss-dis_impl_sub}

\begin{figure}
\resizebox{\hsize}{!}{\includegraphics{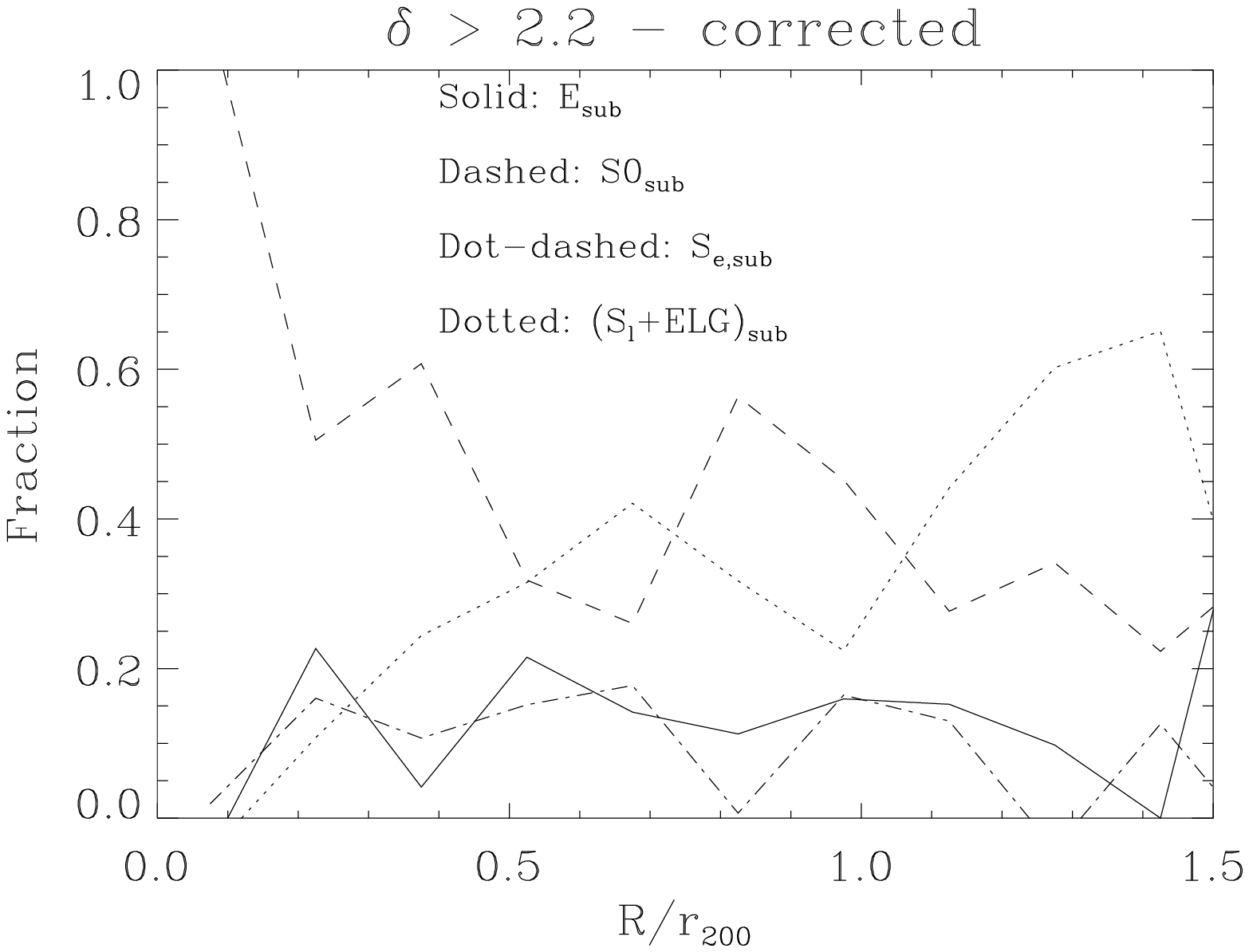}}
\caption{The composition of substructure as a function of 
projected radius $R$.}
\label{f-subcomp}
\end{figure}

The properties of substructure contain a clear clue about the
evolution of the clusters as a whole, and of the galaxies within
them. The fraction of galaxies in substructure, which on average is
about 0.22 (not to be confused with the canonical value of one-third
for the fraction of clusters with significant substructure, e.g.
Geller \& Beers \cite{gb82}; Dressler \& Schectman \cite{ds88}; Jones
\& Forman \cite{jf92}) is correlated with the amount of central
concentration of the total galaxy population. While the central
concentration increases with time, the fraction of galaxies in
substructure decreases. In other words: substructure is destroyed in
the course of time, probably mostly through tidal stripping in the
central regions, on time-scales that are fairly short. The longest
survival times are for high-density groups on non-radial orbits (e.g.,
Gonz\'{a}lez-Casado et al. \cite{go94}).

In this picture, the strong decrease of the fraction of galaxies in
substructure for $R \la 0.3 \, r_{200}$ is not surprising. However,
the new and unexpected result is that the composition of the
substructure changes towards the centre, and in particular within $R
\sim 0.3 \, r_{200}$. Most noticeably, the fraction of S0's in
substructure increases strongly towards the centre. This is evident
from inspection of Fig.~\ref{f-rv8}; however, in view of the
$R$-distribution of the S0's outside substructure, one wonders how
much of this effect could be due to contamination by a small fraction
of S0's outside substructure (the largest galaxy class in our
clusters).

In Fig.~\ref{f-subcomp} we show the composition of substructure as a
function of $R/r_{200}$. The relative contributions of the 4 classes
have been corrected for accidental substructure. This was done by
subtracting from the observed numbers the average number of galaxies
of each type found in 100 azimuthal rescramblings of the data. I.e.,
whatever contamination is still present in Fig.~\ref{f-rv8} has been
taken out in Fig.~\ref{f-subcomp}. The latter figure thus confirms the
reality of the increase of the fraction of S0's, which is accompanied
by a decrease of the fraction of S$_l$+ELG towards the centre in
substructure. In other words: it appears that in the substructure that
survives in the central regions, the S$_l$+ELG's have a harder time to
survive than the S0's. On the basis of the evidence in
Fig.~\ref{f-subcomp} it thus seems that the late spirals in subclumps
are not protected very well against stripping and 'conversion'. The
fractions of E's and S$_e$'s in substructure do not show a dependence
on distance, but this may partly due to limited statistics.

As a matter of fact, the fraction of S$_l$+ELG's in substructure
within $R/r_{200} < 0.3$ ($=0.04 \pm 0.02$) is even smaller than that
of the S$_l$+ELG outside subtructure within $R/r_{200} < 0.3$ ($0.19
\pm 0.04$), at the 3.8 $\sigma$- level. This probably implies that,
even within subclumps, late spirals suffer stripping through
ram-pressure and turbulence or viscosity that is similar to the
general stripping held responsible for the transformation of spirals
into S0's (e.g. Abadi et al. \cite{ab99}; Quilis \& Moore
\cite{qu01}). Actually, the data seem to imply that the tidal
stripping of the subclumps, when they approach the centre, makes life
even harder for the late spirals in substructure than for the late
spirals outside substructure. However, the processes by which the life
of late spirals becomes harder when the subclumps in which they find
themselves approach the centre are not at all clear, as transformation
within subclumps of late spirals into S0's does not seem very likely.

\subsection{Galaxies outside substructure}
\label{ss-dis_impl_nosub}

For the galaxies outside substructure (remember, this is the large
majority), the situation is different. This population contains a
contribution of galaxies (especially in the central region) that
entered the cluster as members of subclumps. However, it is likely
that most of the galaxies outside substructure entered the cluster
from the field, mostly as late-type galaxies, and partly as E's that
formed in small groups prior to entry into the cluster (Pence
\cite{pe76}; Dressler \cite{dr80a}; Merritt \cite{me85}). This infall 
picture is also supported by evidence on the radial anisotropy of the
orbits of the late-type galaxies (as seen e.g. in the ENACS cluster
sample, see Papers III and VI).

Spurred by the seminal results of Butcher \& Oemler (\cite{bo84}) and
Dressler et al. (\cite{dr97}) there have, in recent years, been many
investigations, both observational and theoretical, of the influence
of the cluster environment on the galaxies. The morphological
composition of clusters is observed to change with redshift, and the
usual interpretation is that S0's have formed relatively recently
through transformation of spirals (Dressler et al. \cite{dr97}; Fasano
et al. \cite{fa00}). The details of the transformation were studied in
several ways. Poggianti et al. (\cite{po99}) discuss spectral evidence
for a two-stage transformation, while Jones et al. (\cite{jo00})
conclude that the most likely progenitors of S0's in rich clusters are
early spirals in which starformation was quenched.

The mechanisms by which galaxies in clusters may evolve and transform
have been modelled by several groups, e.g. Moore et al. (\cite{mo98},
\cite{mo99}), Abadi et al. (\cite{ab99}), Diaferio et al. (\cite{di00}),
and Okamoto \& Nagashima (\cite{on01}). The processes that can affect
the morphologies of discs are e.g. stripping of gas through
ram-pressure (Gunn \& Gott \cite{gg72}; Abadi et al. \cite{ab99}) or
turbulence and viscosity (Quilis et al. \cite{qu01}), impulsive tidal
interactions between galaxies (Moore et al. \cite{mo98}, \cite{mo99}),
and mergers (e.g. Barnes \& Hernquist \cite{bh96}). Comparison of the
results of such models with observations is not trivial, because model
parameters must be translated into observables. E.g., deriving the
morphological type from a bulge-to-disk ratio alone may not be subtle
enough, and that could be one of the reasons why Okamoto \& Nagashima
(\cite{on01}) fail to reproduce the well-established
morphology-density relation.

>From numerical models, Moore et al. (\cite{mo99}) found that the fate of
spiral galaxies in clusters depends very much on the amount of central
concentration of the distribution of total mass. Spirals with slowly
rising rotation curves (i.e. not centrally concentrated) have between
50 and 90\% of their stars stripped after 10 Gyrs or so, through
impulsive interactions with other galaxies. On the contrary, the
galaxies with more centrally concentrated mass distributions can
survive relatively unscathed, albeit that the scale height of their
stellar disc in general increases through tidal heating.

In the context of our segregation results, it is quite relevant that
there is a significant correlation between the morphology of a spiral
galaxy and the form of its rotation-curve (see e.g. Corradi \&
Capaccioli \cite{cc90}; Biviano et al. \cite{bi91}; Adami et
al. \cite{ad99}; Dale et al. \cite{da01}). Flat rotation curves (which
indicate a centrally peaked mass distribution) are seen much more
often for early spirals than for late spirals, for which the rotation
curves are more often rising (indicating a less centrally peaked mass
distribution). The relative paucity of late spirals in the central
region probably indicates that most late spirals in the centre have
been destroyed. On the other hand, the early spirals can survive in
the centre, and the different radial distributions of early and late
spirals probably just reflect the different shapes of their potential
wells.

As shown in Paper VIII, the fraction of ELG among early spirals is
significantly lower than the fraction of ELG among late spirals ($0.19
\pm 0.03$ and $0.56 \pm 0.05$, respectively, for the sample considered
in the present paper). Part of this difference could be intrinsic
(Gavazzi et al. \cite{ga98}), but there may also be a contribution
from the difference in the radial distribution. In any case, the
result is similar to the radial dependence of the HI deficiency as
discussed by Solanes et al. (\cite{so01}). Actually, within 1.0 Abell
radius ($\sim 1.25 \, r_{200}$) the HI deficiency is systematically
higher for the early spirals than it is for the late spirals. Solanes
et al. (\cite{so01}) interpret this as evidence for the fact that
early spirals are more easily emptied of their gas. The difference in
radial distribution could also be a factor, although Thomas \& Katgert
(\cite{tk02}) find no evidence for that.

Finally, we turn to the relation between the early spirals and the
S0's. While their $(R,v)$-distributions are different according to the
KS2D comparison, closer inspection reveals that most of the signal for
that difference is in relative velocity. Actually, the radial
distribution of the early spirals is indistinguishable from that of
the S0's. This may or may not be good news, depending on one's
prejudices about the radial variation of the efficiency with which
impulsive encounters transform early spirals into S0's. Yet, the data
seem to indicate an almost constant efficiency of the transformation
of early spirals into S0's. If this is indeed the case, the density
increase towards the centre apparently is largely offset by the larger
velocities. In this picture, the difference in line-of-sight velocity
dispersion of the early spirals and S0's, within $\sim 0.3 \,
r_{200}$, must have a natural explanation. E.g., it could be that the
early spirals that have survived have obtained a velocity distribution
that makes them relatively insusceptible to transformation into an S0.

\section{Summary and conclusions}
\label{s-summ}

We have studied evidence for luminosity and morphology segregation in
an ensemble cluster of $\sim 3000$ galaxies with positions,
magnitudes, velocities, and galaxy type, in clusters observed in the
ESO Nearby Abell Cluster Survey. From positions and velocities we
identify galaxies in and outside substructure. The fraction of
galaxies in substructure appears to decrease strongly towards the
cluster centre. Luminosity segregation is evident only for the very
bright ($M_R \la -22$) ellipticals outside substructure, which mostly
are brightest cluster members near the centres of their clusters.

For galaxies of all types, we find that those within substructure are
segregated with respect to those outside substructure. This is mostly
due to the fact that galaxies in and outside substructure have very
different radial distributions. In addition, morphology segregation is
found among galaxies both {\em in} and {\em outside} substructure.

The early- and late-type galaxies {\em outside substructure} have
different $(R,v)$-distributions, i.e. of projected position $R$ and
relative velocity $v$. The early-type galaxies except the brightest
ellipticals all have very similar $(R,v)$-distributions, i.e. the
fainter ellipticals and S0's are not segregated. Similarly, the late
spirals and the emission-line galaxies have indistinguishable
$(R,v)$-distributions, but the $(R,v)$-distributions of the early
spirals differs from that of the early-type galaxies and from that of
the other late-type galaxies. Among galaxies {\em in substructure},
the S0's are segregated from the late spirals and the emission-line
galaxies, separately as well as together.

Luminosity segregation is most likely due to the dissipative processes
in the innermost region of the clusters, which presumably produce the
brightest ellipticals. The decrease of the fraction of galaxies in
substructure towards the centre is probably due to tidal disruption.
The cause for the accompanying decrease of the fraction of late-type
galaxies in the subclumps is not evident, in particular because the
latter is even stronger than the decrease towards the centre of the
fraction of late-type galaxies outside substructure. 

The large difference between the radial distributions of early and
late spirals are attributed to systematic differences in their mass
profiles. The late spirals presumably are fairly easily destroyed
through impulsive encounters with other galaxies, and the early
spirals much less, so that they can survive in the inner cluster
regions. We briefly discuss the constraints that our data provide for
the process by which early spirals transform into S0's.

\begin{acknowledgements}
AB and PK acknowledge the hospitality of Leiden Observatory and
Trieste Observatory, respectively. This research was partly supported
by the {\em Consorzio Nazionale per lo studio della Formazione ed
evoluzione delle galassie} and by the Leids Kerkhoven-Bosscha
Fonds. We thank Tim de Zeeuw for a careful reading of the manuscript.

\end{acknowledgements}

\vspace*{1.0cm}

\appendix

\section{The clusters used in the analysis}
\label{s-samples}

Details about the 59 clusters used in the analysis are given in
Table~\ref{t-systems}, which contains the following information:

\noindent Column 1: The ACO number of the system \\
	  Columns 2 and 3: The adopted position of the cluster centre \\
          Column 4: Average velocity of the system in the CMBR
	  reference frame (km~s$^{-1}$) \\
	  Column 5: The overall velocity dispersion computed from all 
          cluster members (km~s$^{-1}$) \\
	  Column 6: Total number of galaxies with redshifts, N$_z$ \\ 
	  Column 7: Number of member galaxies with ENACS redshifts, N$_m$ \\ 
	  Column 8: Number of member galaxies with ENACS redshifts and
          galaxy type, N$_t$ \\

Note that the two systems in A548 with essentially the same redshift
are spatially distinct; the one at 12278 km~s$^{-1}$ is the eastern component
of this double cluster, that at 12736 km~s$^{-1}$ is the western component.
For 17 clusters we had no CCD-imaging (ACO nrs. 524, 1809, 2048,
2819 ($\overline{v}=22285$), 2819 ($\overline{v}=25864$), 3151, 3158,
3194, 3202, 3365, 3651, 3691, 3705, 3822, 3827, 3921 and 4010), and in
6 clusters (ACO nrs. 2799, 3112, 3122, 3559, 3825 and 3827) the
fraction of galaxy types from CCD-imaging is less than 50\%.

\begin{table*}
\caption[]{The 59 ENACS clusters with at least 20 member galaxies, and
           galaxy types for at least 80\% of the members}
\begin{center}
\begin{tabular}{||r|r@{\hspace{0.2cm}}r|rr|rrr||r|r@{\hspace{0.2cm}}r|rr|rrr||}
\hline
ACO  & $\alpha_{centre}$ & $\delta_{centre}$ & $\overline{v}_{3K}$ & $\sigma_p$ & N$_z$ & N$_m$ & N$_t$ & ACO  & $\alpha_{centre}$ & $\delta_{centre}$ & $\overline{v}_{3K}$ & $\sigma_p$ & N$_z$ & N$_m$ & N$_t$ \\
 & \multicolumn{2}{c|}{1950.0} & & & & & & & \multicolumn{2}{c|}{1950.0} & & & & & \\
\hline
  13 & 00 11 00 & -19 46.7 & 27949 &  897 &  37 &  37 &  37 & 3094 & 03 09 48 & -27 09.9 & 20027 &  654 &  66 &  66 &  64 \\
  87 & 00 40 13 & -10 04.3 & 16149 &  875 &  27 &  27 &  27 & 3111 & 03 15 55 & -45 51.8 & 23179 &  770 &  35 &  35 &  34 \\
 119 & 00 53 40 & -01 30.3 & 12997 &  720 & 104 & 102 &  87 & 3112 & 03 16 12 & -44 25.2 & 22417 &  954 &  76 &  67 &  60 \\
 151 & 01 06 27 & -16 12.7 & 12074 &  403 &  25 &  25 &  25 & 3122 & 03 20 21 & -41 31.4 & 19171 &  782 &  90 &  89 &  88 \\
 151 & 01 06 24 & -15 41.2 & 15679 &  747 &  46 &  44 &  44 & 3128 & 03 28 50 & -52 42.9 & 17931 &  765 & 155 & 152 & 152 \\
 151 & 01 06 08 & -15 53.3 & 29459 &  804 &  34 &  33 &  33 & 3151 & 03 38 21 & -28 50.2 & 20352 &  752 &  34 &  34 &  34 \\
 168 & 01 12 35 &  00 05.4 & 13201 &  518 &  76 &  76 &  71 & 3158 & 03 41 25 & -53 47.9 & 17698 & 1006 & 105 & 105 & 102 \\
 295 & 01 59 44 & -01 22.1 & 12490 &  298 &  30 &  30 &  26 & 3194 & 03 57 10 & -30 18.7 & 29111 &  797 &  32 &  32 &  32 \\
 514 & 04 46 11 & -20 34.4 & 21374 &  875 &  90 &  82 &  74 & 3202 & 03 59 24 & -53 49.3 & 20741 &  435 &  27 &  27 &  27 \\
 524 & 04 55 40 & -19 47.0 & 23262 &  802 &  26 &  26 &  25 & 3223 & 04 06 34 & -30 57.2 & 17970 &  597 &  68 &  66 &  65 \\
 548 & 05 46 39 & -25 27.8 & 12400 &  710 & 111 & 108 & 108 & 3341 & 05 23 43 & -31 38.6 & 11364 &  561 &  63 &  63 &  63 \\
 548 & 05 43 26 & -25 55.5 & 12638 &  824 & 125 & 120 & 116 & 3354 & 05 33 04 & -28 34.2 & 17589 &  367 &  58 &  56 &  56 \\
 754 & 09 06 35 & -09 27.1 & 16754 & 1010 &  38 &  38 &  38 & 3365 & 05 46 14 & -21 56.5 & 27879 & 1151 &  32 &  32 &  32 \\
 957 & 10 11 17 & -00 39.3 & 13661 &  649 &  34 &  32 &  31 & 3528 & 12 51 40 & -28 44.4 & 16377 &  971 &  28 &  28 &  28 \\
 978 & 10 17 56 & -06 16.5 & 16648 &  497 &  58 &  56 &  52 & 3558 & 13 25 09 & -31 13.7 & 14571 & 1035 &  75 &  73 &  73 \\
1069 & 10 37 18 & -08 24.8 & 19909 &  937 &  32 &  32 &  31 & 3559 & 13 27 04 & -29 15.4 & 14313 &  425 &  39 &  38 &  37 \\
1809 & 13 50 35 &  05 24.2 & 24169 &  774 &  30 &  30 &  30 & 3562 & 13 31 01 & -31 25.3 & 14633 &  903 & 111 & 105 & 105 \\
2040 & 15 10 21 &  07 36.7 & 13974 &  675 &  37 &  37 &  37 & 3651 & 19 48 10 & -55 11.4 & 17863 &  662 &  79 &  78 &  78 \\
2048 & 15 12 50 &  04 34.1 & 29303 &  661 &  25 &  25 &  25 & 3667 & 20 08 24 & -56 57.8 & 16620 & 1037 & 103 & 103 & 102 \\
2052 & 15 14 15 &  07 11.1 & 10638 &  719 &  33 &  33 &  27 & 3691 & 20 30 55 & -38 12.7 & 26021 &  699 &  31 &  31 &  31 \\
2361 & 21 36 08 & -14 32.3 & 17924 &  332 &  24 &  24 &  24 & 3705 & 20 38 54 & -35 23.9 & 26687 & 1059 &  29 &  29 &  29 \\
2401 & 21 55 36 & -20 20.6 & 16844 &  475 &  23 &  23 &  22 & 3764 & 21 22 48 & -34 56.9 & 22442 &  583 &  36 &  36 &  34 \\
2569 & 23 14 54 & -13 05.7 & 23897 &  482 &  36 &  36 &  36 & 3806 & 21 42 55 & -57 31.0 & 22825 &  808 &  97 &  84 &  83 \\
2734 & 00 08 47 & -29 08.1 & 18217 &  579 &  83 &  77 &  77 & 3822 & 21 50 40 & -58 06.2 & 22606 &  871 &  84 &  84 &  68 \\
2799 & 00 35 02 & -39 24.3 & 18724 &  423 &  36 &  36 &  36 & 3825 & 21 54 44 & -60 41.5 & 22373 &  699 &  59 &  59 &  57 \\
2800 & 00 35 29 & -25 20.9 & 18777 &  400 &  33 &  33 &  27 & 3827 & 21 58 26 & -60 10.8 & 29338 & 1132 &  20 &  20 &  20 \\
2819 & 00 43 46 & -63 49.0 & 22285 &  409 &  49 &  49 &  44 & 3879 & 22 24 05 & -69 16.7 & 19982 &  444 &  42 &  42 &  36 \\
2819 & 00 43 54 & -63 52.2 & 25864 &  353 &  43 &  43 &  41 & 3921 & 22 46 43 & -64 41.7 & 27907 &  495 &  31 &  30 &  30 \\
2911 & 01 23 51 & -38 13.5 & 24012 &  404 &  29 &  27 &  27 & 4010 & 23 28 34 & -36 46.7 & 28437 &  622 &  30 &  30 &  30 \\
3093 & 03 09 15 & -47 35.1 & 24771 &  408 &  22 &  21 &  20 &      &            &           &       &      &     &     &     \\ 
\hline
\end{tabular}					 
\label{t-systems}
\end{center}
\end{table*}

\section{KS2D tests with ambiguous results, involving galaxies in substructure}
\label{s-ambiv}

Here we give details about the 12 KS2D comparisons which involve at
least one sample of galaxies inside substructure, and which do not
give identical results at all 4 values of $\delta_{lim}$ (see
Sect.~\ref{ss-segrules}). In Table~\ref{t-sub_amb} the results of
each of these 4 comparisons are given, where an 'X' indicates that the
probability that the two samples are drawn from the same parent sample
is less than 5\%. We have interpreted the results as follows.

In general, there are two effects that can mask a real difference:
limited statistics, and contamination by galaxies outside substructure
in a sample of galaxies within substructure. Imagine a comparison
between a sample of galaxies outside substructure and a sample of
galaxies in substructure. The former will hardly be contaminated by
galaxies in substructure, but the latter will inevitably contain some
galaxies outside substructure. These contaminating galaxies may mask a
real difference, namely when their distribution is similar to that of
the galaxies in the other sample. On the other hand, contamination can
also induce a spurious difference between two samples. If the galaxies
outside substructure, which contaminate the substructure sample, are
distributed differently from the galaxies in the other sample, they
may create a spurious difference. Contamination is more important at
low $\delta_{lim}$, but if we use a higher $\delta_{lim}$, to reduce
contamination, we pay in terms of limited statistics.

\begin{table}
\caption[]{Tests with ambiguous results involving substructure samples}
\begin{flushleft}
\begin{tabular}{|l|c|c|c|c|l|}
\hline
\multicolumn{1}{|c|}{samples} & \multicolumn{4}{c|}{$\delta_{lim}$} & 
\multicolumn{1}{|c|}{adopted} \\
\cline{2-5}
        & 1.8 & 2.0 & 2.2 & 2.4 & \\
\hline
S0$_{sub}$ --- ELG$_{sub}$    & X & X & X &   & different \\ 
S$_{l,nosub}$ --- ELG$_{sub}$ & X &   &   & X & different \\
E$_{nosub}$ --- E$_{sub}$     &   & X & X & X & different \\
E$_{sub}$ --- S0$_{nosub}$    &   & X & X & X & different \\
S$_{e,nosub}$ --- S$_{l,sub}$ &   & X & X & X & different \\
S$_{e,nosub}$ --- E$_{sub}$   &   &   & X & X & different \\
E$_{sub}$ --- ELG$_{sub}$     & X &   &   &   & not different \\
ELG$_{nosub}$ --- S0$_{sub}$  & X &   &   &   & not different \\
S$_{l,nosub}$ --- S0$_{sub}$  & X &   &   &   & not different \\
S$_{l,sub}$ --- ELG$_{nosub}$ &   &   & X &   & different \\
S$_{l,sub}$ --- S0$_{sub}$    &   & X &   &   & different \\
S$_{e,nosub}$ --- S0$_{sub}$  &   & X &   &   & not different \\
\hline
\end{tabular}					 
\end{flushleft}
\label{t-sub_amb}
\end{table}				 

For the comparisons in Table~\ref{t-sub_amb} we now summarize the
reason for our adopted result. In the comparison S0$_{sub}$ vs.
ELG$_{nosub}$ limited statistics is assumed to be cause for the result
at the highest value of $\delta_{lim}$. Limited statistics is also
held responsible for absence of a difference in two of the four
comparisons between S$_{l,nosub}$ and ELG$_{sub}$. In this case, the
difference at $\delta_{lim}=1.8$ is taken to be real, because the
contaminants into the ELG$_{sub}$ sample can only reduce the
significance of the difference (since S$_{l,nosub}$ and ELG$_{nosub}$
share the same distribution). The same explanation probably applies to
the next four comparisons: (E$_{nosub}$ vs. E$_{sub}$, E$_{sub}$
vs. S0$_{nosub}$, S$_{e,nosub}$ vs. S$_{l,sub}$, S$_{e,nosub}$
vs. E$_{sub}$). The contaminants in the 'substructure' samples
probably mask an intrinsically real difference when $\delta_{lim}$ is
set too low.  The opposite effect is considered to determine the
results for the comparisons E$_{sub}$ vs. ELG$_{sub}$, ELG$_{nosub}$
vs. S0$_{sub}$, and S$_{l,nosub}$ vs. S0$_{sub}$, which show a
'significantly different' result only at the lowest
$\delta_{lim}$. Yet, in these cases, contamination of non-substructure
galaxies into the substructure samples, is likely to be responsible
for the difference.

The interpretation of the last three comparisons is less clear.  In
the S$_{l,sub}$ vs. ELG$_{nosub}$ comparison, the sample of
S$_{l,sub}$ is very small, and we certainly run into problems of
limited statistics at high $\delta_{lim}$ values. On the other hand,
at low $\delta_{lim}$, a significant contamination of S$_l$ outside
substructure can make the distribution of the S$_{l,sub}$ sample
resemble that of ELG$_{nosub}$. Therefore, we conclude that these
samples are probably different. The four comparisons between
S$_{l,sub}$ and S0$_{sub}$ only give one significantly different
result. However, given the small number of S$_{l,sub}$, this is
remarkable. We therefore think we can trust this result, and ascribe
the other negative results to a problem of limited statistics. In the
last comparison, statistics are much less of a problem, and we should
expect the result at $\delta_{lim}=2.0$ to be confirmed at higher
values of $\delta_{lim}$. As this is not the case, we conclude that
S$_{e,nosub}$ and S0$_{sub}$ are not different.


\vfill
\end{document}